\documentclass[twocolumn]{aastex631}

\usepackage{xspace}
\usepackage{amsmath}
\usepackage{rotating}
\usepackage{multirow}

\newcommand{\prospector}{{\sc Prospector}\xspace}

\begin{document}

\title{JADES: Rest-frame UV-to-NIR Size Evolution of Massive Quiescent Galaxies from Redshift $z=5$ to $z=0.5$ }

\correspondingauthor{Zhiyuan Ji}
\email{zhiyuanji@arizona.edu}

\author[0000-0001-7673-2257]{Zhiyuan Ji}
\affiliation{Steward Observatory, University of Arizona, 933 N. Cherry Avenue, Tucson, AZ 85721, USA}

\author[0000-0003-2919-7495]{Christina C. Williams}
\affiliation{NSF’s National Optical-Infrared Astronomy Research Laboratory, 950 N. Cherry Avenue, Tucson, AZ 85719, USA}
\affiliation{Steward Observatory, University of Arizona, 933 N. Cherry Avenue, Tucson, AZ 85721, USA}

\author[0000-0002-1714-1905]{Katherine A. Suess}
\affiliation{Department of Astronomy and Astrophysics, University of California, Santa Cruz, 1156 High Street, Santa Cruz, CA 95064, USA}
\affiliation{Kavli Institute for Particle Astrophysics and Cosmology and Department of Physics, Stanford University, Stanford, CA 94305, USA}

\author[0000-0002-8224-4505]{Sandro Tacchella}
\affiliation{Kavli Institute for Cosmology, University of Cambridge, Madingley Road, Cambridge, CB3 0HA, UK}
\affiliation{Cavendish Laboratory, University of Cambridge, 19 JJ Thomson Avenue, Cambridge, CB3 0HE, UK}

\author[0000-0002-9280-7594]{Benjamin D.\ Johnson}
\affiliation{Center for Astrophysics $|$ Harvard \& Smithsonian, 60 Garden St., Cambridge MA 02138 USA}

\author[0000-0002-4271-0364]{Brant Robertson}
\affiliation{Department of Astronomy and Astrophysics, University of California, Santa Cruz, 1156 High Street, Santa Cruz, CA 95064, USA}

\author[0000-0002-8909-8782]{Stacey Alberts}
\affiliation{Steward Observatory, University of Arizona, 933 N. Cherry Avenue, Tucson, AZ 85721, USA}

\author[0000-0003-0215-1104]{William M.\ Baker}
\affiliation{Kavli Institute for Cosmology, University of Cambridge, Madingley Road, Cambridge, CB3 0HA, UK}
\affiliation{Cavendish Laboratory, University of Cambridge, 19 JJ Thomson Avenue, Cambridge, CB3 0HE, UK}

\author[0000-0002-4735-8224]{Stefi Baum}
\affiliation{2Department of Physics and Astronomy, University of Manitoba, Winnipeg, MB R3T 2N2, Canada}

\author[0000-0003-0883-2226]{Rachana Bhatawdekar}
\affiliation{
European Space Agency (ESA), European Space Astronomy Centre (ESAC), Camino Bajo del Castillo s/n, 28692 Villanueva de la Cañada, Madrid, Spain}

\author[0000-0001-8470-7094]{Nina Bonaventura}
\affiliation{Steward Observatory, University of Arizona, 933 N. Cherry Avenue, Tucson, AZ 85721, USA}

\author[0000-0003-4109-304X]{Kristan Boyett}
\affiliation{School of Physics, University of Melbourne, Parkville 3010, VIC, Australia}
\affiliation{ARC Centre of Excellence for All Sky Astrophysics in 3 Dimensions (ASTRO 3D), Australia}

\author[0000-0002-8651-9879]{Andrew J.\ Bunker}
\affiliation{Department of Physics, University of Oxford, Denys Wilkinson Building, Keble Road, Oxford OX1 3RH, UK}

\author[0000-0002-6719-380X]{Stefano Carniani}
\affiliation{Scuola Normale Superiore, Piazza dei Cavalieri 7, I-56126 Pisa, Italy}

\author[0000-0003-3458-2275]{Stephane Charlot}
\affiliation{Sorbonne Universit\'e, CNRS, UMR 7095, Institut d'Astrophysique de Paris, 98 bis bd Arago, 75014 Paris, France}

\author[0000-0002-2178-5471]{Zuyi Chen}
\affiliation{Steward Observatory, University of Arizona, 933 N. Cherry Avenue, Tucson, AZ 85721, USA}

\author[0000-0002-7636-0534]{Jacopo Chevallard}
\affiliation{Department of Physics, University of Oxford, Denys Wilkinson Building, Keble Road, Oxford OX1 3RH, UK}

\author[0000-0002-9551-0534]{Emma Curtis-Lake}
\affiliation{Centre for Astrophysics Research, Department of Physics, Astronomy and Mathematics, University of Hertfordshire, Hatfield AL10 9AB, UK}

\author[0000-0003-2388-8172]{Francesco D'Eugenio}
\affiliation{Kavli Institute for Cosmology, University of Cambridge, Madingley Road, Cambridge, CB3 0HA, UK}
\affiliation{Cavendish Laboratory, University of Cambridge, 19 JJ Thomson Avenue, Cambridge, CB3 0HE, UK}

\author[0000-0002-2380-9801]{Anna de Graaff}
\affiliation{Max-Planck-Institut f\"ur Astronomie, K\"onigstuhl 17, D-69117, Heidelberg, Germany}

\author[0000-0002-4781-9078]{Christa DeCoursey}
\affiliation{Steward Observatory, University of Arizona, 933 N. Cherry Avenue, Tucson, AZ 85721, USA}

\author[0000-0003-1344-9475]{Eiichi Egami}
\affiliation{Steward Observatory, University of Arizona, 933 N. Cherry Avenue, Tucson, AZ 85721, USA}

\author[0000-0002-2929-3121]{Daniel J.\ Eisenstein}
\affiliation{Center for Astrophysics $|$ Harvard \& Smithsonian, 60 Garden St., Cambridge MA 02138 USA}

\author[0000-0003-4565-8239]{Kevin Hainline}
\affiliation{Steward Observatory, University of Arizona, 933 N. Cherry Avenue, Tucson, AZ 85721, USA}

\author[0000-0002-8543-761X]{Ryan Hausen}
\affiliation{Department of Physics and Astronomy, The Johns Hopkins University, 3400 N. Charles St., Baltimore, MD 21218}

\author[0000-0003-4337-6211]{Jakob M.\ Helton}
\affiliation{Steward Observatory, University of Arizona, 933 N. Cherry Avenue, Tucson, AZ 85721, USA}

\author[0000-0002-3642-2446]{Tobias J.\ Looser}
\affiliation{Kavli Institute for Cosmology, University of Cambridge, Madingley Road, Cambridge, CB3 0HA, UK}
\affiliation{Cavendish Laboratory, University of Cambridge, 19 JJ Thomson Avenue, Cambridge, CB3 0HE, UK}

\author[0000-0002-6221-1829]{Jianwei Lyu}
\affiliation{Steward Observatory, University of Arizona, 933 N. Cherry Avenue, Tucson, AZ 85721, USA}

\author[0000-0002-4985-3819]{Roberto Maiolino}
\affiliation{Kavli Institute for Cosmology, University of Cambridge, Madingley Road, Cambridge, CB3 0HA, UK}
\affiliation{Cavendish Laboratory, University of Cambridge, 19 JJ Thomson Avenue, Cambridge, CB3 0HE, UK}
\affiliation{Department of Physics and Astronomy, University College London, Gower Street, London WC1E 6BT, UK}

\author[0000-0003-0695-4414]{Michael V.\ Maseda}
\affiliation{Department of Astronomy, University of Wisconsin-Madison, 475 N. Charter St., Madison, WI 53706 USA}

\author[0000-0002-7524-374X]{Erica Nelson}
\affiliation{Department for Astrophysical and Planetary Science, University of Colorado, Boulder, CO 80309, USA}

\author[0000-0003-2303-6519]{George Rieke}
\affiliation{Steward Observatory and Dept of Planetary Sciences, University of Arizona 933 N. Cherry Avenue Tucson AZ 85721, USA}

\author[0000-0002-7893-6170]{Marcia Rieke}
\affiliation{Steward Observatory, University of Arizona, 933 N. Cherry Avenue, Tucson, AZ 85721, USA}

\author[0000-0003-4996-9069]{Hans-Walter Rix}
\affiliation{Max-Planck-Institut f\""ur Astronomie, 
K\""onigstuhl 17, D-69117, 
Heidelberg, Germany}

\author[0000-0001-9276-7062]{Lester Sandles}
\affiliation{Kavli Institute for Cosmology, University of Cambridge, Madingley Road, Cambridge, CB3 0HA, UK}
\affiliation{Cavendish Laboratory, University of Cambridge, 19 JJ Thomson Avenue, Cambridge, CB3 0HE, UK}

\author[0000-0002-4622-6617]{Fengwu Sun}
\affiliation{Steward Observatory, University of Arizona, 933 N. Cherry Avenue, Tucson, AZ 85721, USA}

\author[0000-0003-4891-0794]{Hannah \"Ubler}
\affiliation{Kavli Institute for Cosmology, University of Cambridge, Madingley Road, Cambridge, CB3 0HA, UK}
\affiliation{Cavendish Laboratory, University of Cambridge, 19 JJ Thomson Avenue, Cambridge, CB3 0HE, UK}
\affiliation{Max-Planck-Institut für extraterrestrische Physik, Gießenbachstraße 1, 85748 Garching, Germany}

\author[0000-0001-9262-9997]{Christopher N.\ A.\ Willmer}
\affiliation{Steward Observatory, University of Arizona, 933 N. Cherry Avenue, Tucson, AZ 85721, USA}

\author[0000-0002-4201-7367]{Chris Willott}
\affiliation{NRC Herzberg, 5071 West Saanich Rd, Victoria, BC V9E 2E7, Canada}

\author[0000-0002-7595-121X]{Joris Witstok}
\affiliation{Kavli Institute for Cosmology, University of Cambridge, Madingley Road, Cambridge, CB3 0HA, UK}
\affiliation{Cavendish Laboratory, University of Cambridge, 19 JJ Thomson Avenue, Cambridge, CB3 0HE, UK}

\begin{abstract}

We present the UV-to-NIR size evolution of a sample of 161 quiescent galaxies with $M_* > 10^{10}M_\sun$ over $0.5<z<5$. With deep multi-band NIRCam images in GOODS-South from JADES, we measure the effective radii ($R_e$) of the galaxies at rest-frame 0.3, 0.5 and 1$\micron$. On average, we find that quiescent galaxies are 45\% (15\%) more compact at rest-frame 1$\micron$ than they are at 0.3$\micron$ (0.5$\micron$). Regardless of wavelengths, the $R_e$ of quiescent galaxies strongly evolves with redshift, and this evolution depends on stellar mass.  For lower-mass quiescent galaxies with $M_* = 10^{10}-10^{10.6}M_\sun$, the evolution follows $R_e\propto(1+z)^{-1.1}$, whereas it becomes steeper, following $R_e\propto(1+z)^{-1.7}$, for higher-mass quiescent galaxies with $M_* > 10^{10.6}M_\sun$. To constrain the physical mechanisms driving the apparent size evolution, we study the relationship between $R_e$ and the formation redshift ($z_{\rm{form}}$) of quiescent galaxies. For lower-mass quiescent galaxies, this relationship is broadly consistent with $R_e\propto(1+z_{\rm{form}})^{-1}$, in line with the expectation of the progenitor effect. For higher-mass quiescent galaxies, the relationship between $R_e$ and $z_{\rm{form}}$ depends on stellar age. Older quiescent galaxies have a steeper relationship between $R_e$ and $z_{\rm{form}}$ than that expected from the progenitor effect alone, suggesting that mergers and/or post-quenching continuous gas accretion drive additional size growth in very massive systems. We find that the $z>3$ quiescent galaxies in our sample are very compact, with mass surface densities $\Sigma_e \gtrsim 10^{10} M_\sun/\rm{kpc}^2$, and their $R_e$ are possibly even smaller than anticipated from the size evolution measured for lower-redshift quiescent galaxies. Finally, we take a close look at the structure of GS-9209, one of the earliest confirmed massive quiescent galaxies at $z_{\rm{spec}} \sim 4.7$. From UV to NIR, GS-9209 becomes increasingly compact, and its light profile becomes more spheroidal, showing that the color gradient is already present in this earliest massive quiescent galaxy.

\end{abstract}

\keywords{Galaxy formation(595); Galaxy evolution(594); Galaxy structure(622); High-redshift galaxies(734)}

\section{Introduction} \label{sec:intro}

The origin and evolution of massive and quiescent galaxies that no longer form stars are major open questions in Astrophysics. The last two decades of research have revealed that massive quiescent galaxies emerge rapidly, but in small numbers, at least as early as $z\gtrsim4$. By $z<1$, such galaxies dominate the total stellar mass budget of the Universe \citep[e.g.][]{Muzzin2013, Ilbert2013}. Understanding their evolutionary history is thus a critical piece of building a coherent picture of cosmic evolution.

Galaxy sizes encode information about the spatial distribution of different components within galaxies, including stellar populations, dust, and gas, and how these components evolve. Because size evolution is influenced by both internal processes, such as star formation and feedback, and external ones, such as interactions and mergers, it provides an integrated record of the mechanisms that assemble and transform galaxies.

The near-infrared (NIR) cameras on the Hubble Space Telescope (HST) enabled measurement of the rest-optical structure of massive galaxies up to $z\sim2.5$. A key revelation was the extremely compact morphology of quiescent galaxies at Cosmic Noon: They are five times smaller on average at $z\sim1-2$ compared to their presumed descendants at $z\sim0$ \citep[][just to name a few]{Daddi2005,Trujillo2007,Toft2007,vanDokkum2008,Damjanov2011,Newman2012}, and their stellar mass densities rival the densest stellar structures known in the Universe \citep[e.g.][]{Hopkins2010b}. In addition, compared to star-forming galaxies of similar mass and redshift, quiescent galaxies display more dramatic evolution in size from $z\sim2.5$ to the present \citep[e.g.,][]{Shen2003,vanderWel2014,Barro2017}. Which processes are driving this strong size evolution of quiescent galaxies, and whether the dominant processes change over time have sparked contentious debate. It is expected for galaxies to increase in size over time given the $\Lambda$CDM cosmology, because cosmic mass density decreases over time due to the expansion of the Universe, and galaxy size should, to some extent, reflect the density of the Universe at its formation epoch, i.e. the progenitor effect \citep[e.g.,][]{Peebles1993,Mo2010,Lilly2016}. Disentangling whether the size increase towards lower redshifts is due to the progenitor effect, or that the galaxies themselves increase in size without appreciable increase in mass (e.g. through gas-poor minor mergers) has been a key target of numerous studies \citep[e.g.,][]{Bezanson2009,Hopkins2009,Naab2009,vanDokkum2010,Fan2010,Oser2012,Newman2012,Cassata2013,Williams2017,Suess2019b,Ji2022}. 

How the structure of quiescent galaxies evolves in earlier times -- at $z>2.5$ -- still eludes us. At this redshift range, HST is limited to the wavelengths of rest-frame UV, i.e. bluewards $4000$\AA. Likely because the rest-UV sizes are not representative of the true distributions of stellar mass in galaxies due to non-negligible impact of color gradients \citep[e.g.,][]{Suess2019,Suess2019b,Mosleh2020,Miller2022}, earlier studies relying on HST have found conflicting results: Some studies say that early quiescent galaxy sizes are consistent with extrapolations from lower-redshift quiescent galaxies, while some argue that early quiescent galaxies are exceedingly compact \citep[][]{Straatman2015,Kubo2018,Lustig2021}. Obviously, in the HST era, the major limitation preventing a complete picture was the lack of deep and high resolution rest-frame optical/NIR imaging at $z>2.5$, which prevented both robust photometric characterization due to large uncertainties in Spectral Energy Distribution (SED) modeling, and rest-frame optical morphologies.

The recent launch of JWST \citep{Gardner2023} has opened a new window to view early quiescent galaxies in exquisite detail. Its unprecedented sensitivity and wavelength sampling enable more robust modeling of stellar populations and star formation history (SFH) reconstruction. The arrival of Cycle 1 JWST data demonstrated the power of this new facility to identify and characterize robust massive quiescent galaxy candidates at $z>2.5$, \citep{Carnall2023b,Valentino2023} and make spectroscopic confirmations \citep{Nanayakkara2022, Carnall2023, Glazebrook2023} even as early as $z\sim5$. Of particular importance to understanding their formation pathways, JWST, {\it for the first time}, now allows measurements of their rest-frame optical and NIR structures at high angular resolution and to higher redshifts than previously possible (out to $z\sim10$, well beyond the limit of HST at $z\sim2.5$). 

A quantitative understanding of how quiescent galaxies grow in size across cosmic time is crucial, because structural evolution is directly linked to their formation pathways. The extremely compact morphologies and high stellar mass densities of early quiescent galaxies imply that they assembled their stars rapidly in dense environments, yet by the present day quiescent systems have become significantly more extended in the local Universe. Distinguishing whether this apparent growth in size is dominated by the continuous addition of newly quenched, larger galaxies (progenitor effect), or by physical size increase of individual systems through minor mergers and dynamical processes, has fundamental implications for how galaxies quench and evolve. This work contributes to this effort by using ultra-deep HST and JWST imaging to measure the size of massive quiescent galaxies over $0.5<z<5$. The combination of high angular resolution, homogeneous SED fitting, and a mass-complete sample allows us to trace half-light radii at different rest-frame wavelengths with unprecedented precision.

Specifically, in this paper we use the ultra-deep NIRCam imaging from the JWST Advanced Deep Extragalactic Survey (JADES, \citealt{Eisenstein2023}) to investigate the redshift evolution of the half-light sizes of quiescent galaxies, measured at rest-frame UV, optical and NIR wavelengths. We focus in particular on massive galaxies with stellar mass $\log (M_*/M_\sun) > 10$, in order to focus on the formation pathways mostly independent of environmentally-driven quenching, a process which affects primarily lower-mass galaxies \citep[e.g.,][]{Peng2010, Ji2018,Papovich2018}. Throughout this paper, we adopt the AB magnitude system and the $\Lambda$CDM cosmology with \citealt{Planck2020} parameters, i.e., $\Omega_m = 0.315$ and $\rm{h = H_0/(100 km\,s^{-1}\,Mpc^{-1}) = 0.673}$.

\section{Observations}

The sample of quiescent galaxies presented this work is in the GOODS-South \citep{Giavalisco2004} portion of JADES where we have obtained deep JWST/NIRCam imaging in 9 filters, i.e., F090W, F115W, F150W, F200W, F277W, F335M, F356W, F410M and F444W. At the time of writing, the JADES program is still ongoing, and the parts of observations (PID: 1180 and 1210) included in this study cover a total sky area of $\approx$ 60 arcmin$^2$, including regions around the Hubble Ultra-Deep Field where additional 5 medium-band NIRCam imaging observations, i.e., F182M, F210M, F430M, F460M and F480M, have been obtained by the JWST Extragalactic Medium-band Survey (JEMS, \citealt{Williams2023}, PID: 1963).

We reduce all these NIRCam imaging observations using a consistent method, which has been presented in detail in \citet{Rieke2023}. In short, we process the raw images using  the JWST Calibration Pipeline with some custom corrections. Specifically, during the Stage 1 of the JWST pipeline we mask and correct for the ``snowballs'' effect caused by charge deposition arising from cosmic ray hits. During the Stage 2 of the JWST pipeline we remove the $1/f$ noise and subtract 2D background from the images, and also correct the ``wisp'' features in the short-wavelength (SW, i.e. filters bluer than F277W) images. Afterwards, we tie the astrometry of individual exposures of a given visit to the World Coordinate System (WCS) of a reference catalog constructed from HST/WFC3 F160W mosaics in GOODS-South with astrometry tied to Gaia-EDR3 \citep{GaiaEDR3}. Finally, we combine individual visit-level mosaics together to create the final mosaic, where we choose a pixel scale of 0.03 arcsec pixel$^{-1}$ for all filters. 

We also use the ancillary HST imaging observations taken from the latest Hubble Legacy Fields Data Release in GOODS-South \citep{Illingworth2016, Whitaker2019}, including the imaging data of the ACS/WFC F435W, F606W, F775W, F850LP and F814W filters, and of the WFC3 F105W, F125W, F140W and F160W filters. All these HST images are on the same WCS grid as our NIRCam images.

The JADES photometry is measured using a customized pipeline based on {\sc photutils} (see Section 4 of \citealt{Rieke2023} for details). In brief, source detection was performed using a segmentation map generated from the stacked NIRCam LW image. All imaging data, from HST to NIRCam, were PSF-matched by convolving and resampling each band to a common PSF defined at F444W (see \citealt{Ji2023jems} for details on reconstructing the JADES common PSF models). Fluxes were then measured from the PSF-homogenized mosaics using both segmentation-defined apertures (Kron, isophotal) and a set of circular apertures with different radii, with local background levels estimated and subtracted for each source. In our subsequent SED analysis (Section \ref{sec:sed}), we use the Kron fluxes, which provide the best global measurements of galaxy colors. Photometric uncertainties were calculated from the noise properties of the mosaics and propagated through the measurement process, following methods similar to those in \citet{Labbe2005,Quadri2007}.

\section{Sample Selection} \label{sec:sample}

As the JADES program is still ongoing, the motivation of this work is {\it not} to test, improve or develop selection methods to build a complete demographic picture of quiescent galaxy populations across a wide dynamic range of stellar mass. Instead, we focus this paper only on the population of massive quiescent galaxies with stellar masses $M_*>10^{10}M_\sun$. Because existing HST surveys are remarkably complete at this stellar-mass range ($>90\%$ complete, \citealt{Guo2013,Ji2018}, a detailed discussion regarding sample completeness can be found in Section \ref{sec:final_sample}), our strategy is to make full use of the power of those legacy surveys in GOODS-South to first select candidate galaxies using methods that have been extensively tested in literature, and then refine the selected galaxies using SED fitting with our new JWST data.

\subsection{Selecting Candidate Quiescent Galaxies using Rest-UVJ Colors from 3D-HST}
\label{sec:inital_sample}
Our primary selection of quiescent galaxies is based on the redshift-dependent criteria of rest-frame UVJ colors \citep{Williams2009}. We take advantage of the measurements from the 3D-HST survey \citep{Brammer2012,Skelton2014}. We first impose a cut on the signal-to-noise ratio (S/N $>10$) in the HST/F160W filter, which ensures high-quality photometry and hence those previous measurements of galaxy physical properties from 3D-HST. We then select galaxies with $M_*>10^{9.7}M_\sun$\footnote{The SED fitting code \prospector is elected to use in this work (Section \ref{sec:sed}). It has been noted that the stellar mass measures from \prospector are $0.1-0.3$ dex higher \citep[e.g.,][]{Leja2022,Ji2022}, therefore we use this $10^{9.7}M_\sun$ cut to ensure our initial selection includes the vast majority of galaxies with $M_*\ge10^{10}M_\sun$ from \prospector.} from the 3D-HST catalog. Finally, we use the rest-frame $(U-V)$ and $(V-J)$ colors from the 3D-HST catalog to make the initial selection of candidate quiescent galaxies.

We select candidate quiescent galaxies whose rest-frame colors satisfy the UVJ criteria from \citet{Schreiber2015}:
\begin{equation}
    \begin{split}
    (U-V) > 1.3,\\
    (V-J) < 1.6,  \\ 
    (U-V) > 0.88(V-J) + 0.49.
    \end{split}
    \label{equ:uvj}
\end{equation}
We add to our analysis additional seven\footnote{The most galaxies from \citet{Carnall2020} have already been selected by the UVJ technique described above.} $z>2$ quiescent galaxies from \citet{Carnall2020} which appear in the JADES footprint. \citet{Carnall2020} made the selection of $M_*>10^{10}M_\sun$ quiescent galaxies at $z>2$ based on their refined SED modelling that allows sufficiently flexible solutions over a wide range of observed redshifts and specific star formation rates (see their Section 2.3 and Table 2). \citet{Carnall2020} showed that their selection is highly consistent with the UVJ technique, with a 80\% overlapping fraction, but is more complete in that their SED modelling enabled the selection of relatively young quiescent galaxies which have bluer $(U-V)$ colors that would otherwise have been missed using the lower-redshift UVJ selection criteria.

The above selection gives us a total of 268 candidate massive quiescent galaxies at $z>0.5$. We then remove galaxies hosting bright AGN from our analysis, especially because the presence of AGN can affect the interpretation of their morphological measures. We first remove X-ray AGN by cross matching the sample with the X-ray catalog from the 7Ms Chandra Deep Field South \citep{Luo2017}. We then remove IR AGN using IRAC colors \citep{Donley2012, Ji2022AGN}. We  also remove radio AGNs identified by the latest and deepest VLA radio observations in GOODS-South (Rujopakarn et al. 2023 in preparation). Finally, we remove other types of AGN  identified by recent SED fitting from \citet{Lyu2022}. Adding all these different AGN types together, we identify a total of 71 AGN-hosting galaxies, corresponding to a fraction of $26\pm4\%$ (71/268) which is in broad agreement with that found in galaxies of similar masses at Cosmic Noon \citep[e.g.][]{Xue2010,Ji2022AGN}. We mention that we retain the $z_{\rm{spec}}=4.658$ quiescent galaxy GS-9209 that has recently been spectroscopically confirmed with the JWST/NIRSpec observation by \citet{Carnall2023}. This galaxy is not identified as an AGN based on the selection methods considered above, but it does host a faint AGN, as suggested by the broad H$\alpha$ emission in the NIRSpec spectrum. However, because this AGN is so faint that its broad-band size measurement -- except for the morphology in F356W which covers H$\alpha$ (Section \ref{dis:9209}) -- is not significantly biased by AGN contribution, we decide to keep this galaxy in the following analysis.

After removing the AGN-hosting galaxies, our initial selection based on 3D-HST measurements yields a sample of 197 candidate quiescent galaxies at $0.5<z<5$.

\subsection{Finalizing the Quiescent Sample using JADES Data} \label{sec:final_sample}

\begin{figure*}
    \centering
    \includegraphics[width=0.97\textwidth]{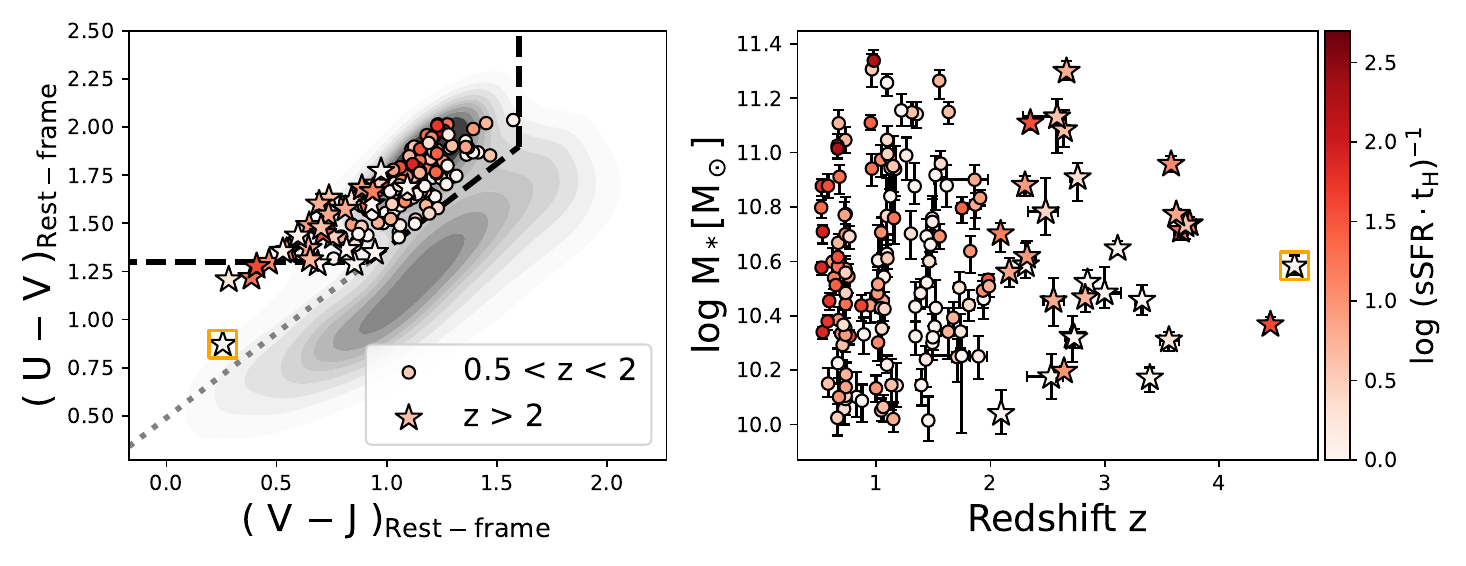}
    \caption{The final sample of 161 massive quiescent galaxies presented in this work (Section \ref{sec:sample}). Quiescent galaxies at $z>2$ are shown as stars, while those at $0.5<z<2$ are shown as circles. The orange square marks the quiescent galaxy GS-9209 at $z_{\rm{spec}}=4.658$ \citep{Carnall2023}. Each one of the galaxies is color-coded according to its star formation intensity, quantified using $(\frac{1}{\rm{sSFR}})/t_{\rm{H}}$ (Section \ref{sec:final_sample}). {\bf Left:} Rest-frame UVJ color diagram. Background grey contours show the distribution of all galaxies with stellar mass $\log M_*>10$ and redshift $0.5<z<5$ from 3D-HST. Black dashed lines show the criteria from \citet{Schreiber2015}, which we use for the selection of quiescent galaxies at $z<2$. Following recent studies, we relax the criterion of the horizontal cut on $(U-V)$ to select quiescent galaxies at $z\ge2$ (Section \ref{sec:sample}). {\bf Right:} The distribution of the quiescent galaxy sample in the stellar mass-redshift space. }
    \label{fig:uvj}
\end{figure*}

We note that the 3D-HST rest-frame colors were computed using EAzY, whose template sets are less comprehensive than those employed in a full SED-fitting framework. We therefore refine the rest-frame color measurements using SED fitting with the new photometry from JADES. We perform SED fitting with a total of more than 20-filter photometry from HST through JWST using the code \prospector (\citealt{Johnson2021}, see Section \ref{sec:sed} for details).  We then use these newly obtained best-fit SEDs to re-measure the rest-frame UVJ colors. At $z<2$, we retain the galaxies still satisfying all three UVJ color cuts (Equation \ref{equ:uvj}).  At higher redshift ($z>2$), most objects still lie within the conventional $UVJ$ box; however, recent studies have shown that a subset of $z\gtrsim 2$ quiescent galaxies can fall just outside the traditional boundary, primarily due to slightly bluer $U-V$ colors \citep[e.g.,][]{Valentino2023}. This is generally interpreted as reflecting younger stellar ages at earlier cosmic times \citep{Whitaker2013,Belli2019,Carnall2020}. Motivated by these results, for $z>2$ we adopt a slightly relaxed criterion by requiring galaxies to satisfy the diagonal color cut (the grey dotted line in Figure \ref{fig:uvj}), following the recommendation of \citet[see their Section 3.1]{Carnall2020}. We did not apply this relaxation in our initial selection because broadening the $UVJ$ region can substantially increase contamination from  star-forming galaxies  \citep{Carnall2020,Valentino2023}. Mitigating this contamination typically requires additional cuts based on SFR or sSFR, which are challenging to constrain robustly for quiescent systems (see later discussion in this section). Since our primary goal is to measure the size evolution of quiescent galaxies, we prioritized sample purity in the selection; the relaxed $z>2$ criterion would add only a small number of sources ($<10\%$ of the $z>2$ sample, \citealt{Carnall2020}, also, see \citealt{Baker2025aa}). Finally, we note that, as shown in Figure \ref{fig:uvj}, only a few galaxies lie outside the conventional UVJ box. We verified that including or excluding them does not change any of our conclusions. 

Our final sample consists of 161 robust UVJ-selected quiescent galaxies with $M_*>10^{10}M_\sun$ at $0.5<z<5$, among which 130 and 31 galaxies are at $z<2$ and $z\ge2$, respectively.  Among the 36 removed galaxies (i.e., 
$197-161$), 18 were excluded because their stellar masses are $<10^{10}M_\odot$ based on our \prospector\ fits (note that our initial mass cut in the 3D-HST catalog is $10^{9.7}M_\odot$ to account for possible systematics in stellar-mass estimates, see Section \ref{sec:inital_sample}). The remaining 18 galaxies were removed because their \prospector-derived UVJ colors fall outside the UVJ selection box. To investigate the origin of this discrepancy, we re-ran their SED fitting with \prospector\ using pre-JWST photometry and found that the vast majority (15/18) still lie outside the UVJ box. This shows that inaccurate UVJ colors from EAzY, likely due to its limited template set, are the main reason for their misclassification.

Recent studies also used alternative selection methods of quiescent galaxies based on a fixed or redshift-dependent threshold of specific star formation rate (sSFR) \citep[e.g.,][]{Speagle2014,Donnari2019,Leja2022,Tacchella2022}. However, accurately measuring sSFRs  for quiescent galaxies with photometry only is challenging  \citep{Conroy2013}, which is highly sensitive to SED model assumptions \citep[e.g.,][]{Pacifici2023}. As Appendix \ref{app:prior} shows, we observe a significant systematic offset in the sSFR measures by simply changing SFHs (Section \ref{sec:sed}), meaning that whatever sSFR thresholds we use to select quiescent galaxies would be highly model-dependent. By contrast, the rest-frame UVJ color measures are largely insensitive to the SED model assumptions, making it a more empirical way to select quiescent galaxies. Because of this, we choose to still use the UVJ technique in this work. Nevertheless, the UVJ selection is highly consistent with the selection methods based on sSFR. To show this, in the left panel of Figure \ref{fig:uvj} we plot the distribution of our final quiescent sample in the rest-frame UVJ diagram. We color code each one of the galaxies according to $(\frac{1}{\rm{sSFR}})/t_{\rm{H}}$, i.e., the ratio of the mass doubling time at the current star formation rate\footnote{Here the current star formation rate is defined as the star formation rate of the first lookback time bin, i.e. 0$-$30 Myr, from our fiducial SED fitting (Section \ref{sec:sed}).} to the Hubble time at the time of observation. A larger $(\frac{1}{\rm{sSFR}})/t_{\rm{H}}$ corresponds to a lower star-formation intensity. Regardless of SFH assumptions (Appendix \ref{app:prior}), we find that the UVJ-selected quiescent galaxies have longer mass doubling time than the Hubble time, confirming that the UVJ-selected quiescent galaxies also have low sSFRs.

The final quiescent sample obtained here is highly mass complete at $z>3$. \citet{Leja2020} estimated the 100\% mass-complete limit for the full 3D-HST survey to be $\sim10^{10.1}M_\odot$ at $z\sim3$ (see their Figure 1). The GOODS-S region considered in this work is deeper than the other 3D-HST fields by $\sim0.2{-}0.5$ mag in HST/F160W, implying that our adopted mass threshold of $10^{10}M_\odot$ is safely about the 100\% mass completeness limit. In addition, massive quiescent galaxies at high redshift are intrinsically bright: at $z>2.5$ our galaxies span $\mathrm{F444W}=21{-}24$~AB, comparable to NIRCam-identified quiescent systems at similar redshifts (e.g., \citealt{Baker2025aa}), and more than five magnitudes brighter than the $5\sigma$ limit of JADES \citep{Rieke2023}. These magnitudes are also $>1$~mag brighter than the pre-JWST IRAC depths in GOODS-S ($\sim25$~AB; \citealt{Skelton2014}), meaning that all galaxies already had robust IRAC detections with high signal-to-noise ($\mathrm{S/N}\gtrsim10$ in both Channel 1 and Channel 2). Previous template-fitting analyses of IRAC photometry in GOODS-S demonstrated excellent consistency among deblending methods at these bright magnitudes \citep{Guo2013}, indicating that rest-frame UVJ colors were already well constrained for these galaxies in the pre-JWST imaging.

Independent indications further support the high completeness of the sample at $z>3$. The number densities of massive quiescent galaxies derived from our sample are fully consistent with those inferred in other JWST legacy fields within uncertainties (\citealt{Baker2025jades}), suggesting that no significant population is missed in our selection. Moreover, using the recently released JADES SED-fitting catalog \citep{Simmonds2025}, we verify that all galaxies at $3<z<5$ with $\log(M/M_\odot)>10$ and $\log(\mathrm{sSFR/yr^{-1}})<-11$ are present in our final sample. Altogether, these lines of evidence demonstrate that the quiescent sample presented here is highly complete down to $10^{10}M_\odot$ over $0.5<z<5$. Any residual incompleteness, if present, is expected to be negligible and should not significantly affect any of the conclusions reached in this work.

\section{Measurements}

\subsection{SED Fitting with Prospector} \label{sec:sed}

With the JADES photometry that samples the wavelength range from rest-frame UV to NIR with more than 20 filters, we perform SED fitting using the code \prospector \citep{Johnson2021}, with the emphasis on robustly measuring stellar mass, photometric redshift and sSFR. We assume a similar SED model used in \citet{Ji2023jems}. In brief, we adopt the FSPS stellar synthesis code \citep{Conroy2009,Conroy2010} with the stellar isochrone libraries MIST \citep{Choi2016,Dotter2016} and the stellar spectral libraries MILES \citep{Falcon-Barroso2011}. We adopt the \citet{Madau1995} IGM absorption model and include the nebular emission model of \citet{Byler2017}. We model the dust attenuation following \citet{Tacchella2022} where the dust attenuation of nebular emission and young stellar populations, and of old stellar populations are treated differently \citep{Charlot2000}. We fix the redshift when a spectroscopic one is available\footnote{We use the spectroscopic redshift catalog in GOODS-South compiled by the ASTRODEEP project \citep{Merlin2021}.}, otherwise we fit it as a free parameter with a flat prior of $z\in(0.1,10)$. 

As the fiducial model of this study, we assume nonparametric, piece-wise SFH composed of 9 lookback time bins, where SFR is constant within each bin. We model nonparametric SFH using the Dirichlet prior \citep{Leja2017} which can recover high-fidelity SFHs of different galaxy types while effectively mitigate the overfitting problem \citep{Leja2019}. In addition, this prior has been demonstrated, with synthetic observations from cosmological simulations, to give the unbiased inference of stellar ages for massive quiescent galaxies at $z\sim2$ \citep[see Appendix A in][]{Ji2022}. 

In Figure \ref{fig:w_3dhst} we compare the redshifts and stellar masses from our \prospector fitting with those from 3D-HST \citep{Brammer2012,Skelton2014}. The agreement between the redshift measures is excellent. Our \prospector measurement of stellar masses is also tightly correlated with that from 3D-HST, although our stellar masses are systematically higher by $\approx 0.16$ dex, a quantitatively similar systematics has also been found by other studies \citep[e.g.,][]{Leja2019,Lower2020,Leja2022,Ji2022,Ji2023}. In Appendix \ref{app:prior}, we further test our measurements with another two different SFHs, namely nonparametric SFHs with the continuity prior \citep{Leja2019} and delayed-tau parametric SFHs. We show that our measurements of stellar mass, rest-frame UVJ colors (hence the selection of quiescent galaxies) are robust.

\begin{figure}
    \centering
    \includegraphics[width=0.47\textwidth]{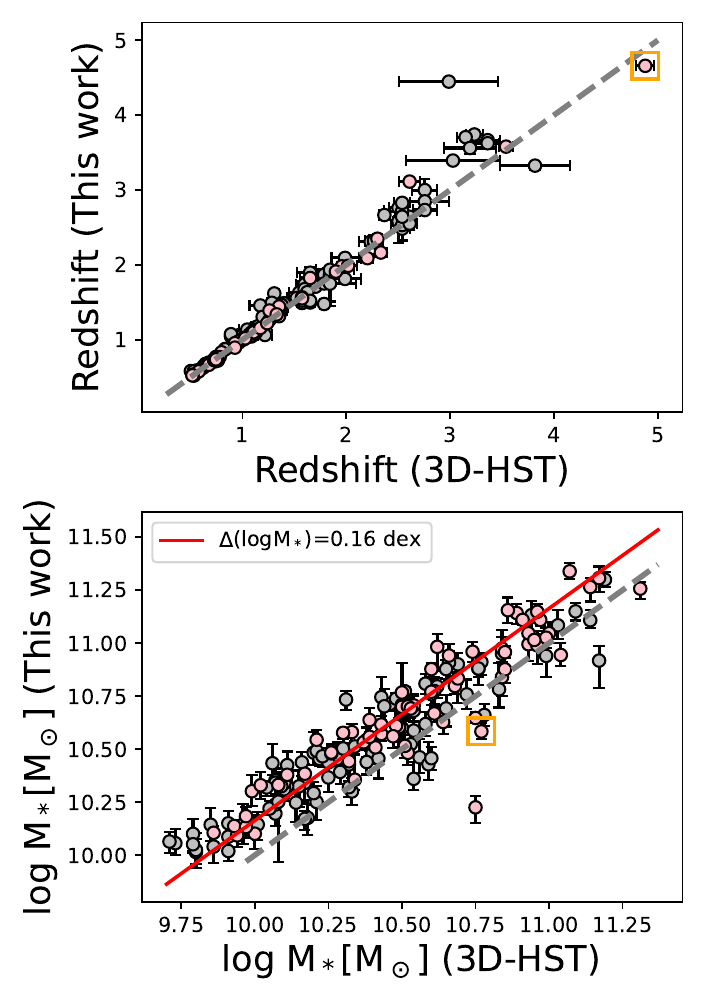}
    \caption{Comparisons of the redshift (top) and stellar-mass (bottom) measures. Galaxies with spectroscopically confirmed redshifts are color coded in pink. GS-9209 is marked with the orange square. The grey, dashed line marks the one-to-one relationship. The uncertainty of photometric redshifts from our \prospector fitting is smaller than the point, hence hard to see in the plot. Our stellar-mass measures with \prospector are systematically higher by 0.16 dex relative to the measures from 3D-HST.} \label{fig:w_3dhst}
\end{figure}

\subsection{Measuring Rest-frame Sizes} \label{sec:galfit}

\begin{figure}
    \centering
    \includegraphics[width=0.447\textwidth]{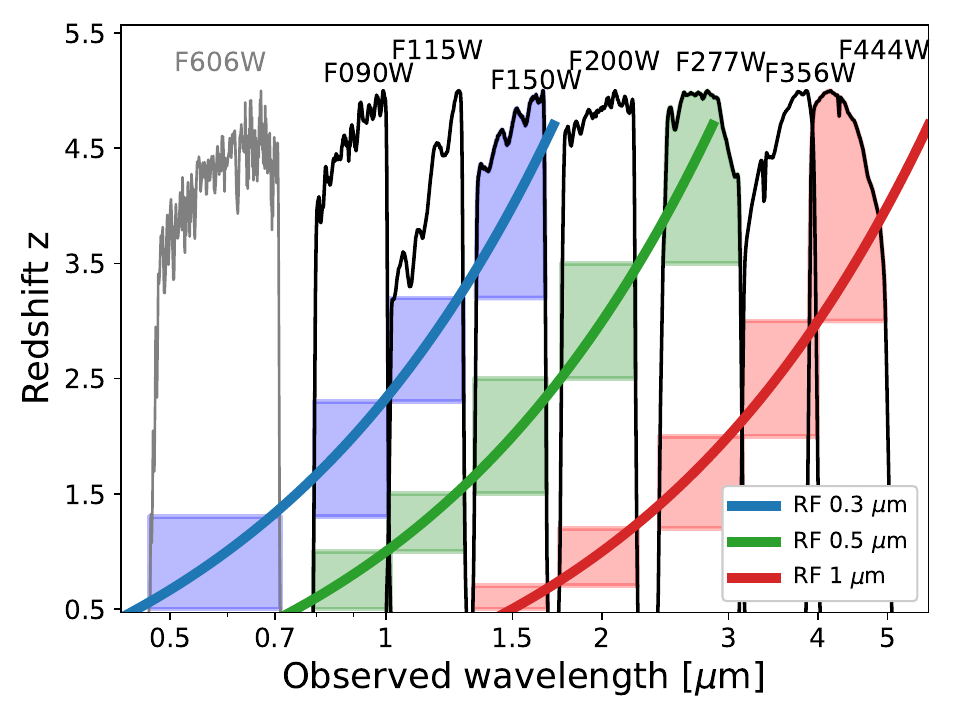}
    \caption{Wide-band filters used for the rest-frame 0.3$\micron$ (blue), 0.5$\micron$ (green) and 1$\micron$ (red) size measures. Expect the HST/ACS F606W, all other filters are from JWST/NIRCam.}
    \label{fig:filter_selection}
\end{figure}

In this paper we refer to the circularized effective (half-light) radius ($R_e = R_{e,maj}\times\sqrt{b/a}$) as the size of galaxies\footnote{We find no substantial changes in our conclusions if we use the effective semi-major axis instead.}. We measure the size at three characteristic rest-frame wavelengths -- 0.3, 0.5 and 1$\micron$ -- using different wide-band images. As Figure \ref{fig:filter_selection} illustrates, we can perform the most measurements using NIRCam images from JADES, except the rest-0.3$\micron$ sizes for galaxies at $z<1.3$ which we instead use the HST/ACS F606W images. We have tested our results by linearly interpolating\footnote{For the two $z>4$ galaxies in our sample, we actually need do the extrapolation to get their rest-frame $1\micron$ sizes.} the measurements of two adjacent filters to get the size at the exact rest-frame wavelength. We found no substantial changes in any of our conclusions.

We measure the size by performing PSF-convolved morphological fitting using the software {\sc Galfit} \citep[version 3.0.5,][]{galfit}, where each one of the galaxies is modelled as a single 2-dimensional S\'{e}rsic profile. Before running {\sc Galfit}, we first center on each one of the galaxies to make a $7''\times7''$ cutout. Similar to previous studies \citep[e.g.,][]{vanderWel2012}, we only allow the S\'{e}rsic index $n$ varying between 0.2 and 8. To mitigate possible impacts from  adjacent galaxies on the morphological fitting, we use the JADES segmentation map \citep{Rieke2023} to mask out all neighboring galaxies in the cutouts. By default, we fit the local sky background of each cutout as a free parameter. The PSFs used for the fitting are the model PSFs (mPSFs) of JADES \citep{Ji2023jems}. The mPSFs are generated using the software {\sc Webbpsf} \citep[v1.1.1,][]{webbpsf}, where we have carefully taken into account the effect of data reduction. As demonstrated by \citet[][see their Appendix A]{Ji2023jems}, the mPSFs are highly consistent with the commonly-used effective PSFs \citep[ePSFs,][]{Anderson2000} to the level of $\approx 1\%$ in terms of enclosed energy radial profiles. The biggest advantage of using mPSFs, especially for the extragalactic deep fields like JADES which contain very limited number of bright, unsaturated isolated stars in their footprints, is that it allows us to have robust, stable PSF models to larger ($>1''$) angular scales.

We estimate uncertainties of the fitted parameters using a Monte Carlo method. Specifically, we use the error flux extension of the NIRCam data to Monte Carlo resample the image pixel values 300 times. We then run {\sc Galfit} on the resampled images and use the range between 16th and 84th percentiles as $1\sigma$ uncertainties. In this paper, because we only focus on massive galaxies which are detected with S/N $>$ 100 in the JWST images\footnote{For the galaxies presented in this paper, the median S/N is 220 in F150W, and is 450 in F444W, where S/N is measured within the Kron aperture.}, the derived uncertainty of $R_{e}$ is generally very small. For reference, we find a median uncertainty of NIRCam/F150W size to be $\delta R_{e}/R_{e}\approx2\%$, which is in quantitative agreement with the uncertainties estimated for galaxies with similar S/Ns using HST/F160W images (see Figure 6 in \citealt{vanderWel2012} and Figure 6 in \citealt{Ji2022AGN}).

\begin{figure}
    \centering
    \includegraphics[width=0.47\textwidth]{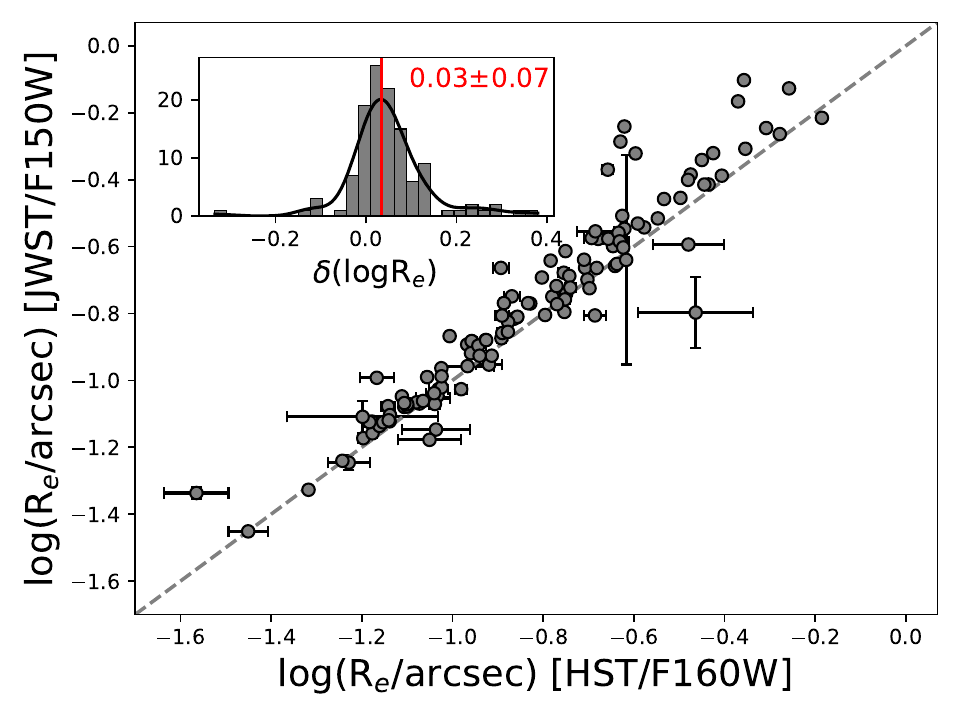}
    \caption{Size from our measurements using the NIRCam/F150W images ($y$-axis, Section \ref{sec:galfit}) as a function of size from \citet{vanderWel2012} who performed the measurements using the HST/WFC3 F160W images ($x$-axis). The agreement between the two measurements is great, with a median difference ($y-x$) of 0.03 dex and a standard deviation of 0.07 dex.}
    \label{fig:f150w_v_f160w}
\end{figure}

We now test systematics of our {\sc Galfit} fitting. First, we consider the systematics introduced by the local sky background estimation. Instead of fitting the sky as a free parameter, if we fix it to the $3\sigma$-clipped median pixel value of the cutout after masking out all detected sources using the segmentation map, the median relative difference in the size measures will be $\approx 1\%$. Second, instead of using mPSFs, if we use ePSFs in the {\sc Galfit} fitting, the median relative difference in the size measures will be $\approx 3\%$. These systematic uncertainties are small and have no impacts on the results of this work. Finally, we compare our NIRCam/F150W sizes to the HST/WFC3 F160W sizes measured by \citet{vanderWel2012}. As Figure \ref{fig:f150w_v_f160w} shows, we find excellent agreement between the two measures, with a median difference of $0.03\pm0.07$ dex. Moreover, among the 161 quiescent galaxies, we find that 24 F150W {\sc Galfit} fits hit the S\'{e}rsic index bounds ($n=0.2$ or 8), which is also consistent with \citet{vanderWel2012} who found 21 F160W fits having the same issue. 

We remove galaxies with $n=0.2$ or 8 from our sample to ensure reliable size measures for the size evolution analysis. Specifically, we remove 37, 25 and 25 galaxies from the analysis of the rest-frame 0.3$\micron$, 0.5$\micron$ and 1$\micron$ size evolution, respectively. We note that all of the removed fits have $n=8$, indicating that those galaxies have very compact cores. Removing them from our analysis thus can drag the median size up. To check the magnitude of this impact, for the galaxies with unreliable single S\'{e}rsic fits, we use the alternative approach with the Richardson–Lucy deconvolution algorithm (see Section \ref{sec:test_re_zgt3} below for details) to estimate their sizes, and then add these alternative sizes back to the analysis. Despite that the best-fit parameters of the size evolution are consistent with those reported in Table \ref{tab:size_evo} within uncertainty, the intercept $C$ (Equation \ref{equ:size}) systematically reduces by 0.08, 0.06 and 0.05 dex for rest-frame 0.3$\micron$, 0.5$\micron$ and 1$\micron$ sizes, respectively. Other than this, we find no substantial changes in any of our results.

\subsubsection{Testing the size measures for the $z>3$ galaxies} \label{sec:test_re_zgt3}

As we will show in Section \ref{sec:size_evo}, the $z>3$ quiescent galaxies in our sample have typical sizes of $R_e^{1\micron}\sim0.3$ kpc, corresponding to 0".04 at $z=3$ which is only slightly larger than the 0".03 pixel scale of JADES mosaics. Moreover, all the $z>3$ quiescent galaxies have large S\'{e}rsic indices $n>2$. Long tails of large S\'{e}rsic index profiles can make the analysis sensitive to profile mismatch and sky background estimates \citep{galfit}. We thus conduct several further tests to check the robustness of the rest-frame 1$\micron$ size estimates for these extremely compact galaxies at $z>3$. 

\begin{figure}
    \centering
    \includegraphics[width=0.47\textwidth]{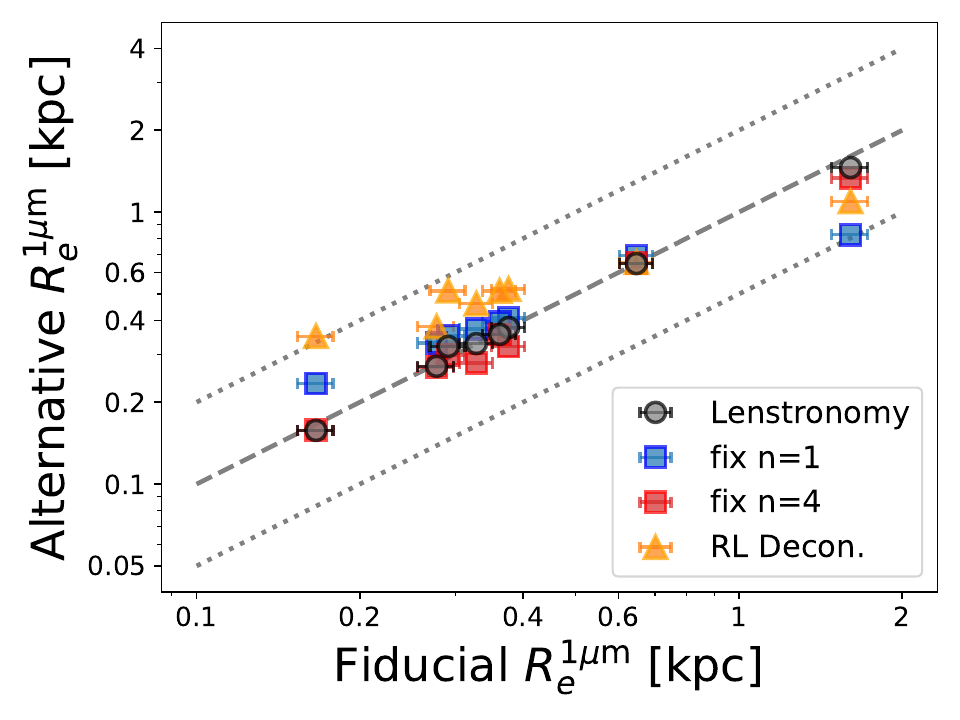}
    \caption{Rest-frame 1$\micron$ sizes of the $z>3$ quiescent galaxies from different methods. The $x$-axis shows the fiducial measurements in this work (Section \ref{sec:galfit}). The $y$-axis shows the measurements from alternative methods detailed in Section \ref{sec:test_re_zgt3}, including using (1) a different morphological fitting tool {\sc Lenstronomy} (black circles), (2) {\sc Galfit} but fixing S\'{e}rsic index $n=1$ (blue squares) or $n=4$ (red squares) during the fit and (3) a nonparametric method with the Richardson–Lucy deconvolution (orange triangles). The main purpose here is to compare the sizes from different methodologies, we thus do not estimate uncertainties for the $y$-axis. The dashed line marks the one-to-one relation, and the dotted lines mark the 2 times above/below the one-to-one relation. For the vast majority of the $z>3$ quiescent galaxies, the relative difference in sizes from different methods is $<50\%$. }
    \label{fig:size_zgt3}
\end{figure}

First, the robustness of S\'{e}rsic fitting, especially for galaxies with very compact morphologies, relies on accurate pixelization and integration of S\'{e}rsic profiles. As shown by \citet[][see their Figure 7]{Robotham2017}, the flux weighted error model image generated by {\sc Galfit} increases with S\'{e}rsic $n$. We thus test our S\'{e}rsic fitting for the $z>3$ quiescent galaxies with {\sc Lenstronomy} (Suess et al., in preparation), a newly developed morphological fitting code that adopts new algorithms to generate S\'{e}rsic models and to conduct fitting. We plot the results as black circles in Figure \ref{fig:size_zgt3}. Excellent agreement is seen between the sizes from {\sc Galfit} and from {\sc Lenstronomy}.

Second, the degeneracy between $R_e$ and S\'{e}rsic $n$  \citep[e.g.][]{Ji2020,Ji2022AGN} makes the size measures uncertain. To check how sensitive the size estimates are due to uncertain $n$, instead of fitting $n$ as a free parameter, we run {\sc Galfit} by fixing $n=1$ (exponential disk) or $n=4$ (de Vaucouleurs profile) for the $z>3$ quiescent galaxies. In Figure \ref{fig:size_zgt3} the results are shown as blue and red squares, respectively. The sizes from the $n=4$ S\'{e}rsic fitting are in very good agreement with our fiducial measurements, which is expected since all the galaxies have $n>2$ from the default S\'{e}rsic fitting. The agreement between the fiducial sizes and those from the $n=1$ fitting is also good, with the differences well within a factor of 2. We note that, relative to our default S\'{e}rsic fitting, the reduced $\chi^2$ increases by a factor of 1.8 and 1.1 for the fitting with fixed $n=1$ and $n=4$, respectively, showing that the quality of fitting is better with $n$ as a free parameter, as expected. 

Third, it is known that the light profile of quiescent galaxies can be more complex than a single S\'{e}rsic profile \citep[e.g][]{Huang2013a, Huang2013b}. \citet{Davari2014} simulated quiescent galaxies at different redshifts by mimicing their light profiles using the detailed substructures observed in nearby elliptical galaxies. They found that the systematic uncertainty of the size measures at $z\sim2$ from single S\'{e}rsic fitting is at the level of $\sim10-20$\%. If high-redshift quiescent galaxies indeed have similarly multi-component light profiles like lower-redshift counterparts, the tests done by \citet{Davari2014} will then suggest that the mismatch of assumed light profiles only introduces subtle bias in the inferred sizes. However, because the rest-frame NIR light distribution of quiescent galaxies at $z>3$ remains largely unknown, we further test our size measures using the nonparametric approach with the Richardson–Lucy deconvolution algorithm \citep{Richardson1972}. In brief, Richardson–Lucy deconvolution is an iterative method for recovering an estimate of a source’s intrinsic image given an observed image blurred by a known PSF. It assumes the image formation is essentially observed
$\approx$ true $\ast$ PSF,
and updates the current estimate of the “true” image by comparing (pixel-by-pixel) the observed image to a PSF-blurred version of the current estimate. The update is multiplicative and is derived as a maximum-likelihood solution under Poisson noise, which naturally enforces non-negative flux. With more iterations it can sharpen structure.

With precise PSF models, images can be restored to some accuracy through the iterative Richardson–Lucy algorithm. As a sanity check, we run Richardson–Lucy deconvolution for the mPSF image (Section \ref{sec:galfit}) of F444W, i.e. the filter used for the rest-1$\micron$ size measures at $z>3$ (Figure \ref{fig:filter_selection}). After $\sim50$ iterations, the FWHM of the deconvolved mPSF image asymptotes to 0".015, i.e. half the pixel size which is the theoretical limit we can achieve for JADES images with the Richardson–Lucy deconvolution, showing the effectiveness of this technique. Thus our strategy is to first deconvolve the F444W images of the $z>3$ quiescent galaxies, and then directly estimate the effective radii ($R_e^{\rm{RL}\dagger}$) of the galaxies by measuring their radial light profiles in the deconvolved images. In this way the PSF effect can be largely mitigated, and the difference between $R_e^{\rm{RL}\dagger}$ of the galaxies and that of the PSF, i.e. 0".015, can be considered as a result of the intrinsic broadening due to the sizes of the galaxies. The sizes can then be inferred following
\begin{equation}
    R_e^{\rm{RL}}[\rm{arcsec}] = \sqrt{(R_e^{\rm{RL}\dagger})^2-(0.015)^2} 
\end{equation}
 where $R_e^{\rm RL\dagger}$ is the half-light radius measured directly from the deconvolved image using the photometric curve of growth computed with circular apertures.

The biggest advantage of the method above based on the Richardson–Lucy deconvolution is that it does not require an assumed form for the parametric light profile, and any complex structures in the galaxies are naturally taken into account during the $R_e^{\rm{RL}}$ estimates. Moreover, this method does not bias towards brighter parts of the galaxies (which differs from {\sc Galfit} which conducts $\chi^2$ minimization fitting and hence essentially provides light-weighted measures). The disadvantage of this method, though, is that no formal stop criterion exists for the iterative process \citep{Prato2012}, as such image deconvolution is mathematically ill-posed. Previous HST studies by \citet{vanderWel2011} fixed the number of iterations to be 16. In our cases, we find that, once the iteration number is greater than 10, the $R_e^{\rm{RL}}$ measures become very stable for the $z>3$ quiescent galaxies.  

In Figure \ref{fig:size_zgt3}, $R_e^{\rm{RL}}$ are shown as orange triangles (based on 50 times Richardson–Lucy iterations). We have visually inspected the deconvolved image for each galaxy and found no clear artifacts introduced by the Richardson–Lucy deconvolution process. Even with this very different approach, we still find very good agreement with the sizes derived from the default single S\'{e}rsic fitting with {\sc Galfit}. We also note that, despite the $z>3$ quiescent galaxies are very compact, our analysis based on the Richardson–Lucy deconvolution suggests that their light profiles are inconsistent with point sources, as all of them have $R_e^{\rm{RL}\dagger}>0".015$, i.e. the galaxies are (marginally) resolved in F444W.

Finally, we stress that the purpose of running different tests here is not to evaluate which methodology is the best in characterizing the morphology of high-redshift quiescent galaxies. In principle, one could use simulated galaxies with assumed light profiles to compare the performance of different size-measurement techniques. However, the scientific usefulness of such tests is fundamentally limited for the present study, because the detailed light distributions of quiescent galaxies at $z>3$ remain poorly constrained. Any simulation-based conclusions would therefore depend on likely over-simplified assumptions about the intrinsic light profiles, which may not capture the true structural diversity of these systems. As a result, such tests could have limited interpretive value and may even introduce misleading biases rather than clarifying the robustness of the measurements. Instead, our goal is to check if the size estimates of the $z>3$ quiescent galaxies suffer from any significant systematic uncertainties from the imperfect fitting algorithm, the degeneracy between $R_e$ and $n$, and the mismatch of assumed light profiles. Regardless of methods, we find strong correlations between the sizes from our fiducial measurements and the sizes from other alternative methods described above. The difference in sizes from different methods are well within a factor of 2. In fact, the relative difference in size is $<50\%$ for the vast majority of the $z>3$ galaxies (Figure \ref{fig:size_zgt3}). We thus conclude that the size estimates for the quiescent galaxies at $z>3$ in our sample are not sensitive to significant systematic uncertainties, hence the strong size evolution of quiescent galaxies towards $z>3$ reported below (Section \ref{sec:size_evo}) is robust.

\section{Results} \label{sec:results}

\begin{figure*}
    \centering
    \includegraphics[width=0.97\textwidth]{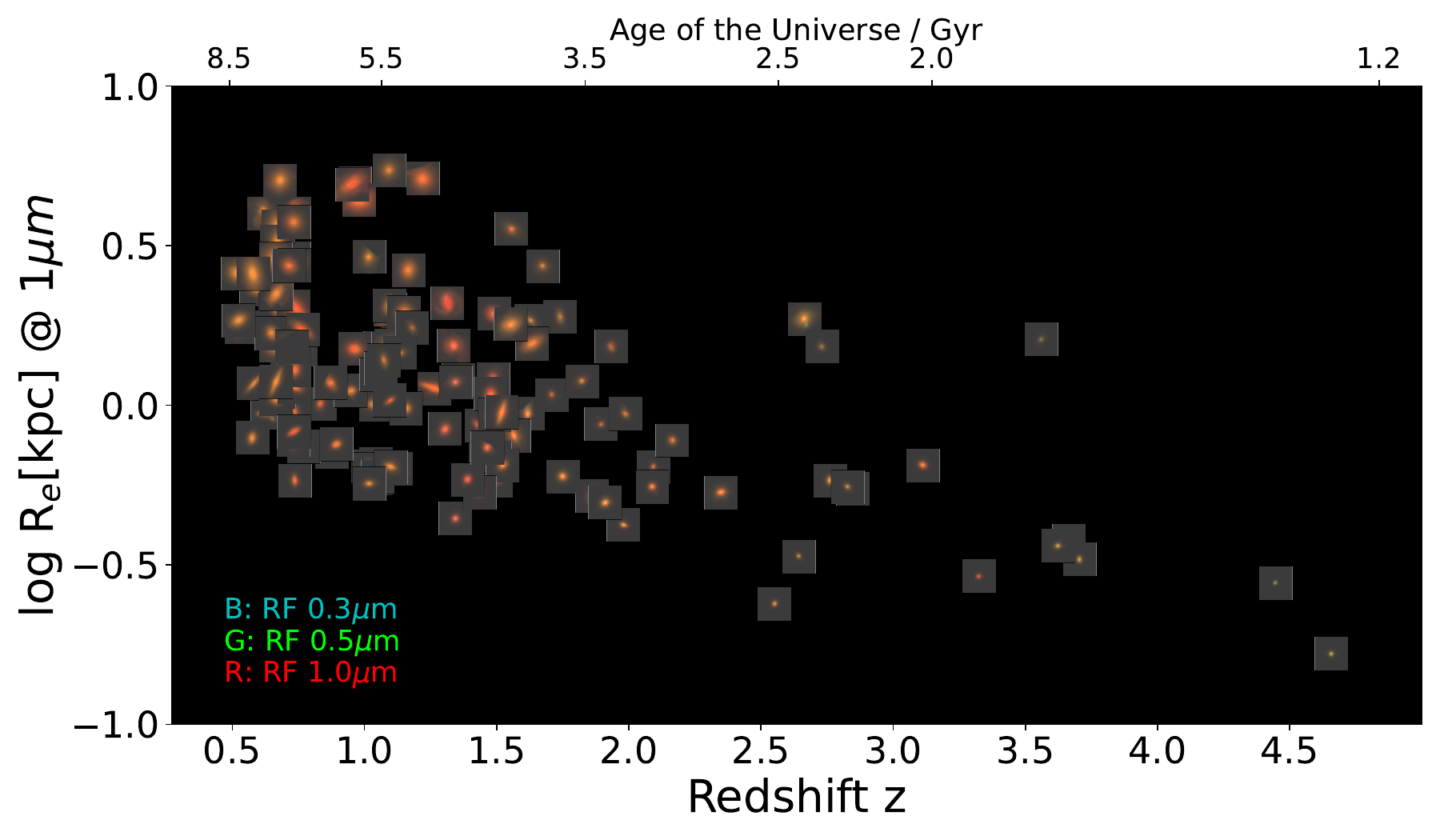}
    \caption{RGB cutouts of massive quiescent galaxies presented in this work. Each one of the galaxies is plotted at its location in the diagram of rest-frame 1$\micron$ size evolution. The filters used to produce the RGB cutouts are illustrated in Figure \ref{fig:filter_selection}. The size of the cutouts is 20 kpc $\times$ 20 kpc.  }
    \label{fig:post}
\end{figure*}

Figure \ref{fig:post} presents the RGB images of individual quiescent galaxies in this work, where the RGB filters are chosen to probe rest-frame 0.3, 0.5, and 1.0$\micron$ (Figure \ref{fig:filter_selection}) and each galaxy is plotted at the corresponding location in the diagram of rest-1$\micron$ size vs redshift. We observe strong size evolution of quiescent galaxies from $z\sim5$ to $z\sim0.5$. In what follows, we will quantify and compare the size evolution at different rest-frame wavelengths from UV through NIR. We will also explore the possible dependence of size evolution on the physical properties of quiescent galaxies. Finally, we will  present the stellar mass-size relationship of massive quiescent galaxies.

\subsection{Multi-wavelength Sizes: Piecing Together the Physical Drivers of Size Evolution} \label{sec:size_proxy}

Because of the spatial variations of dust attenuation and stellar-population properties, galaxies show radial gradients of stellar mass-to-light ratio, which makes the interpretation of the apparent evolution of projected light profiles nontrivial \citep[e.g.][]{Suess2019b,Suess2019,Mosleh2020,Miller2022,vanderWel2023}. In this work, we perform the morphological analysis at three different wavelengths -- that are sensitive to different physical processes -- to constrain the physical drivers of the size evolution of quiescent galaxies. The two shorter wavelengths are chosen to be rest-frame 0.3$\micron$ and 0.5$\micron$, which cover key stellar population features around the rest-frame 4000\AA\ region. The third characteristic wavelength is chosen to be rest-frame 1$\micron$ which is a sensitive probe of stellar-mass distributions, and is the longest wavelength where we can still measure galaxy morphology with the JADES NIRCam imaging over the redshift range considered here (Figure \ref{fig:filter_selection}). 

\begin{figure}
    \centering
    \includegraphics[width=0.47\textwidth]{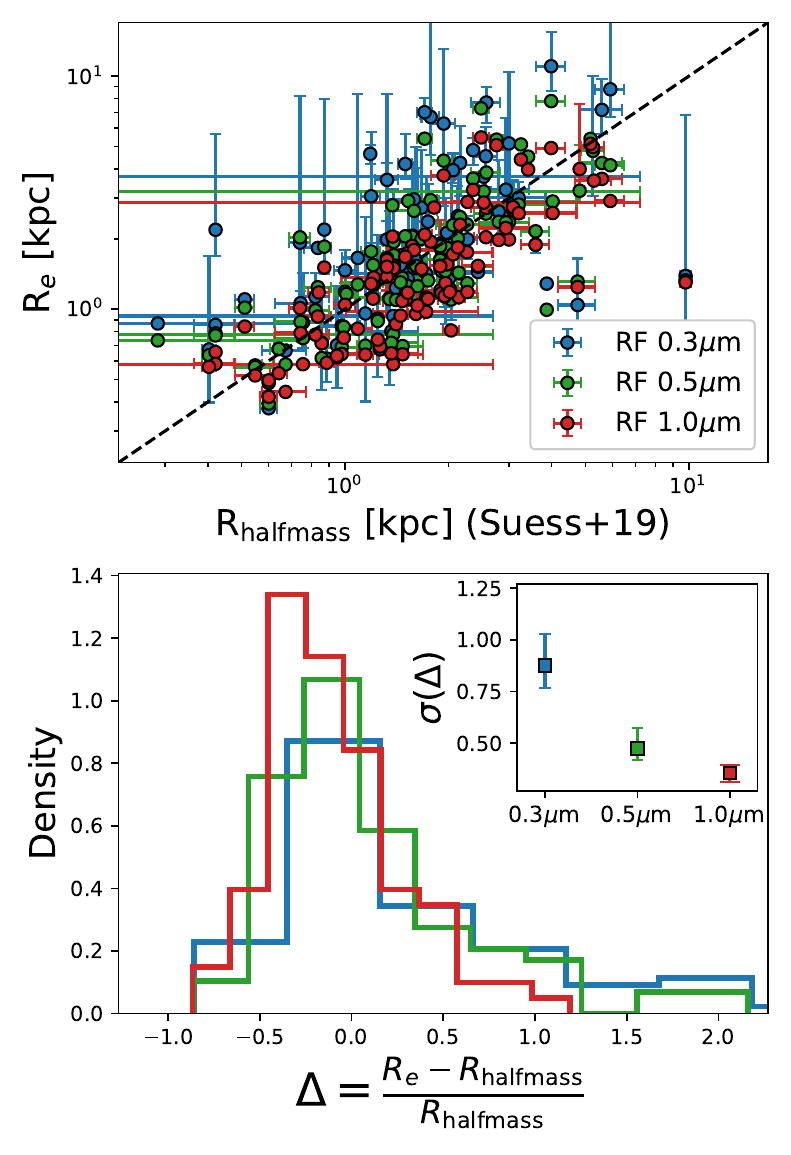}
    \caption{Half-light sizes $R_e$ vs half-mass sizes $R_{\rm{halfmass}}$ for the subsample of quiescent galaxies whose $R_{\rm{halfmass}}$ measures are available from \citet{Suess2019}. Rest-frame 0.3, 0.5 and 1$\micron$ $R_e$ are shown in blue, green and red, respectively. The bottom panel shows histograms of the difference between the half-light and half-mass sizes ($\Delta$), and the inset shows the standard deviation of $\Delta$ (Section \ref{sec:size_evo}). The agreement between $R_{\rm{halfmass}}$ and rest-1$\micron$ $R_e$ is the best. }
    \label{fig:re_rhm}
\end{figure}

Compared to rest-UV and optical, rest-NIR light profiles are a much better proxy for the stellar-mass distribution of galaxies, because the stellar mass-to-light ratio at NIR only weakly, if at all, depends on recent star formation \citep[e.g.,][]{Bell2001,McGaugh2014}. To show this, we cross match our quiescent sample with the catalog from \citet{Suess2020} who measured galaxy half-mass sizes ($R_{\rm{halfmass}}$) using HST data based on spatially resolved SED fitting. A total of 116 galaxies in our sample have the $R_{\rm{halfmass}}$ measures. In the top panel of Figure \ref{fig:re_rhm}, we plot $R_{\rm{halfmass}}$ as a function of $R_e$ measured at different wavelengths. Strong correlations are observed between $R_{\rm{halfmass}}$ and $R_e$, regardless of which wavelength $R_e$ is measured. Systematic offsets are also observed between $R_{\rm{halfmass}}$ and $R_e$, likely due to the different methodologies used for the size measures. 

We compare the agreement between $R_{\rm{halfmass}}$ and $R_e$ measured at different wavelengths. In the bottom panel of Figure \ref{fig:re_rhm}, we show the histograms of $\Delta$, the relative difference between $R_{\rm{halfmass}}$ and $R_e$. We measure the standard deviation of $\Delta$, i.e. $\sigma(\Delta)$. We estimate the uncertainty of $\sigma(\Delta)$ through a bootstrap Monte Carlo method. Specifically, we bootstrap the sample, use normal distributions to Monte Carlo resample the values of $R_{\rm{halfmass}}$ and $R_e$ with the corresponding measurement uncertainties, and finally we measure $\sigma(\Delta)$. We repeat the procedure 1000 times and use the range between 16th and 84th percentiles as the uncertainty of $\sigma(\Delta)$. As the inset of the bottom panel of Figure \ref{fig:re_rhm} shows, the $\Delta$  between $R_{\rm{halfmass}}$ and $R_e^{1\micron}$ has the smallest scatter, showing rest-NIR light profiles indeed a robust, much better proxy for stellar-mass distributions than the rest-UV and optical ones.

In what follows, we thus will use $R_e^{1\micron}$ as the size proxy for stellar-mass distributions of the galaxies. Because the 4000\AA\ break contains key information related to dust attenuation, stellar age and metallicity, combining with the size evolution at rest-frame 0.3$\micron$ and 0.5$\micron$, we will be able to empirically constrain the physical mechanisms driving the size evolution of quiescent galaxies.

\subsection{Size Evolution of Quiescent Galaxies from UV through NIR} \label{sec:size_evo}

\begin{figure*}
    \centering
    \includegraphics[width=0.97\textwidth]{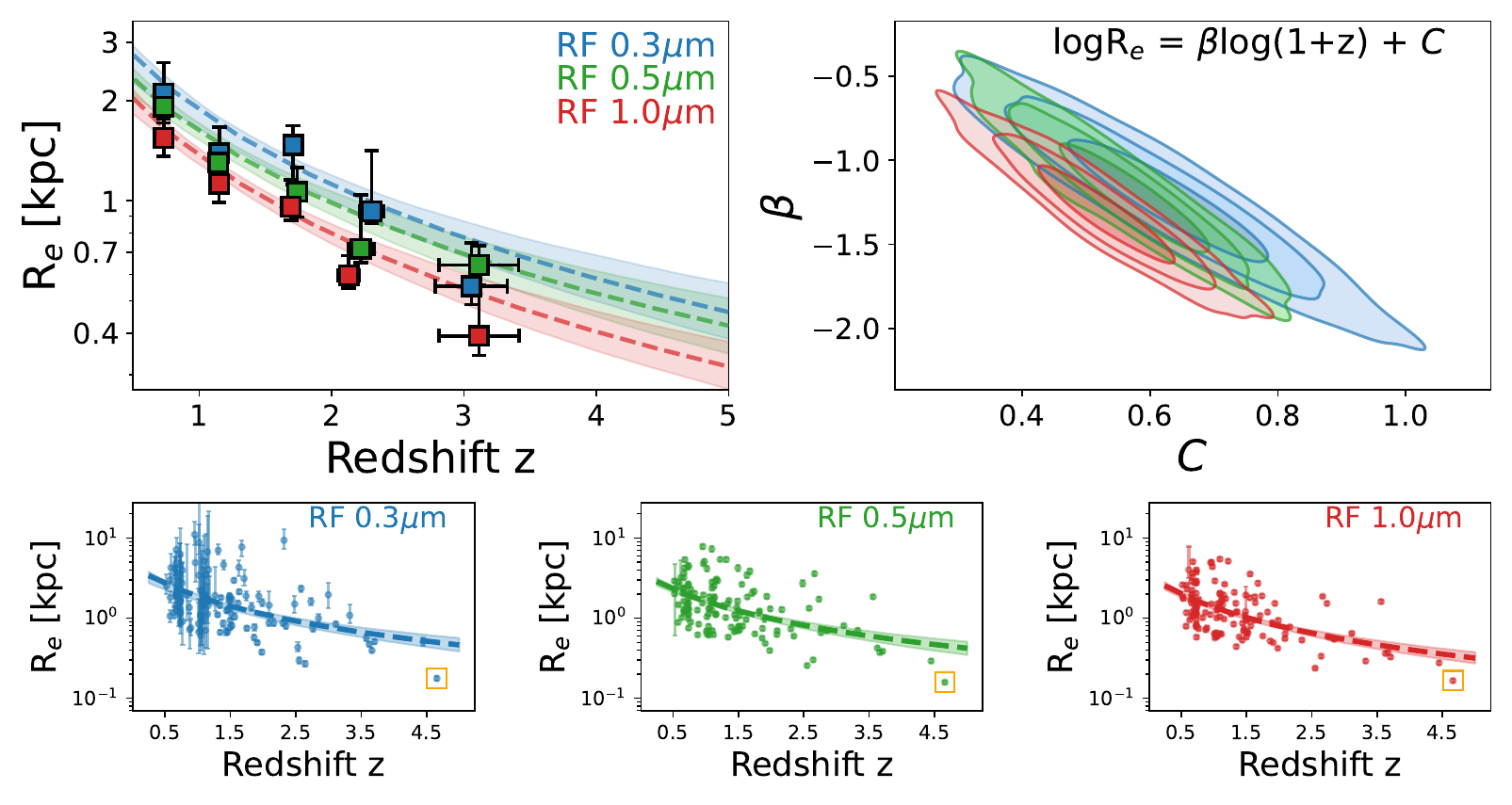}
    \caption{Rest-frame size evolution of massive quiescent galaxies. The corresponding best-fit parameters can be found in Table \ref{tab:size_evo}. {\bf Top:} The left panel shows the best-fit size evolution at rest-frame 0.3$\micron$ (blue), 0.5$\micron$ (green) and 1$\micron$ (red), respectively. The shaded regions mark the $1\sigma$ uncertainties of the best-fit relations. The filled squares with error bars are median $R_e$ and their uncertainties in individual redshift bins (Section \ref{sec:size_evo}). The right panel shows the 1-, 2- and 3-$\sigma$ contours of the $R_e$-z relations based on our bootstrap Monte Carlo method (Section \ref{sec:size_evo}). {\bf Bottom:} Differing from the top-left panel, here we plot individual quiescent galaxies to the size-evolution diagram. Like previous figures, GS-9209 is marked with the orange square.}
    \label{fig:size_evo}
\end{figure*}

We characterize the size evolution by fitting the relationship between the $R_e$ and $z$ of individual quiescent galaxies with a functional form of
\begin{equation}
    \log R_{e} = \beta\log(1+z)+C.
    \label{equ:size}
\end{equation}
To estimate the uncertainties of $\beta$ and $C$, we bootstrap the entire sample, use normal distributions to Monte Carlo resample the size and redshift measures with the corresponding uncertainties, and finally fit the size evolution. We repeat this bootstrap Monte Carlo procedure 1000 times and use the range between 16th and 84th percentiles as $1\sigma$ uncertainties. The best-fit relationships are present in Table \ref{tab:size_evo} and Figure \ref{fig:size_evo}.

\subsubsection{The pace of the size evolution} \label{sec:pace}

Regardless of rest-frame wavelengths, the size of quiescent galaxies evolves with redshift at a similar pace, i.e., a similar $\beta\approx-1.25\pm0.20$ (Table \ref{tab:size_evo}). Because the fitting of size evolution was done using the measurements of individual galaxies and most of the quiescent galaxies in our sample are at $z<2$, the low-redshift quiescent galaxies have more weight in the fitting procedure \footnote{ We verified that the fitted parameters change only subtly when we fit the size evolution using only the 
$z<3$ galaxies in our sample.}. Notwithstanding the very limited sample size at $z>3$, by looking at the bottom panels of Figure \ref{fig:size_evo} we note that the vast majority of $z>3$ quiescent galaxies are below (i.e. have smaller sizes than) the best-fit $R_e$-z relationship. To better show this, we estimate the median sizes of quiescent galaxies at different redshifts. We divide the entire sample into five redshift bins, i.e., four bins between $z=0.5$ and $z=2.5$ with an interval of $\Delta z=0.5$ and the last bin of $z>2.5$. Similarly, we use the bootstrap Monte Carlo method to estimate the uncertainties of the median sizes. As the top-left panel of Figure \ref{fig:size_evo} shows, the median size of the highest redshift bins deviates from the best-fit size evolution -- quiescent galaxies at $z>3$ are smaller than the predictions by extrapolating the size evolution of lower-z quiescent galaxies. We will discuss the implications regarding this finding later in Section \ref{diss:zgt3}.

\begin{table*}[]
    \centering
    \caption{Median sizes of massive quiescent galaxies and their best-fit evolution with redshift.}
    \begin{tabular}{|c | c  c  c  c  c | c c|}
    \toprule
         & $z=0.5\sim1.0$ & $z=1.0\sim1.5$ & $z=1.5\sim2.0$ & $z=2.0\sim2.5$ & $z=2.5\sim5.0$ & $\beta$$^{(b)}$ & $C$$^{(b)}$\\
    \hline
        $R_e(0.3\micron)^{(a)}$ & 2.11$_{-0.33}^{+0.52}$ & 1.39$_{-0.22}^{+0.26}$ & 1.47$_{-0.31}^{+0.21}$ & 0.94$_{-0.08}^{+0.43}$  & 0.56$_{-0.07}^{+0.19}$ &  -1.23$_{-0.24}^{+0.21}$ & 0.62$_{-0.11}^{+0.10}$ \\  
    \hline
        
        $R_e(0.5\micron)^{(a)}$ & 1.90$_{-0.26}^{+0.17}$ & 1.30$_{-0.13}^{+0.16}$ & 1.07$_{-0.14}^{+0.17}$ & 0.72$_{-0.06}^{+0.42}$  & 0.60$_{-0.21}^{+0.11}$ &  -1.24$_{-0.17}^{+0.21}$ & 0.57$_{-0.08}^{+0.07}$ \\
    \hline
        
        $R_e(1.0\micron)^{(a)}$ & 1.55$_{-0.20}^{+0.17}$ & 1.10$_{-0.13}^{+0.13}$ & 0.96$_{-0.09}^{+0.19}$ & 0.6$_{-0.05}^{+0.07}$  & 0.38$_{-0.04}^{+0.17}$ &  -1.33$_{-0.18}^{+0.19}$ & 0.53$_{-0.07}^{+0.06}$\\  
    \hline
    \multicolumn{8}{l}{\small $\dagger$ Values shown in the table are the medians and their 1-$\sigma$ uncertainties from our bootstrap Monte Carlo method (Section \ref{sec:size_evo}).} \\
    \multicolumn{8}{l}{\small (a) In the unit of kpc. (b) The best-fit relation of $\log R_e=\beta\log (1+z) + C$. } \\
    \end{tabular}
    
    \label{tab:size_evo}
\end{table*}

\begin{figure*}
    \centering
    \includegraphics[width=1\textwidth]{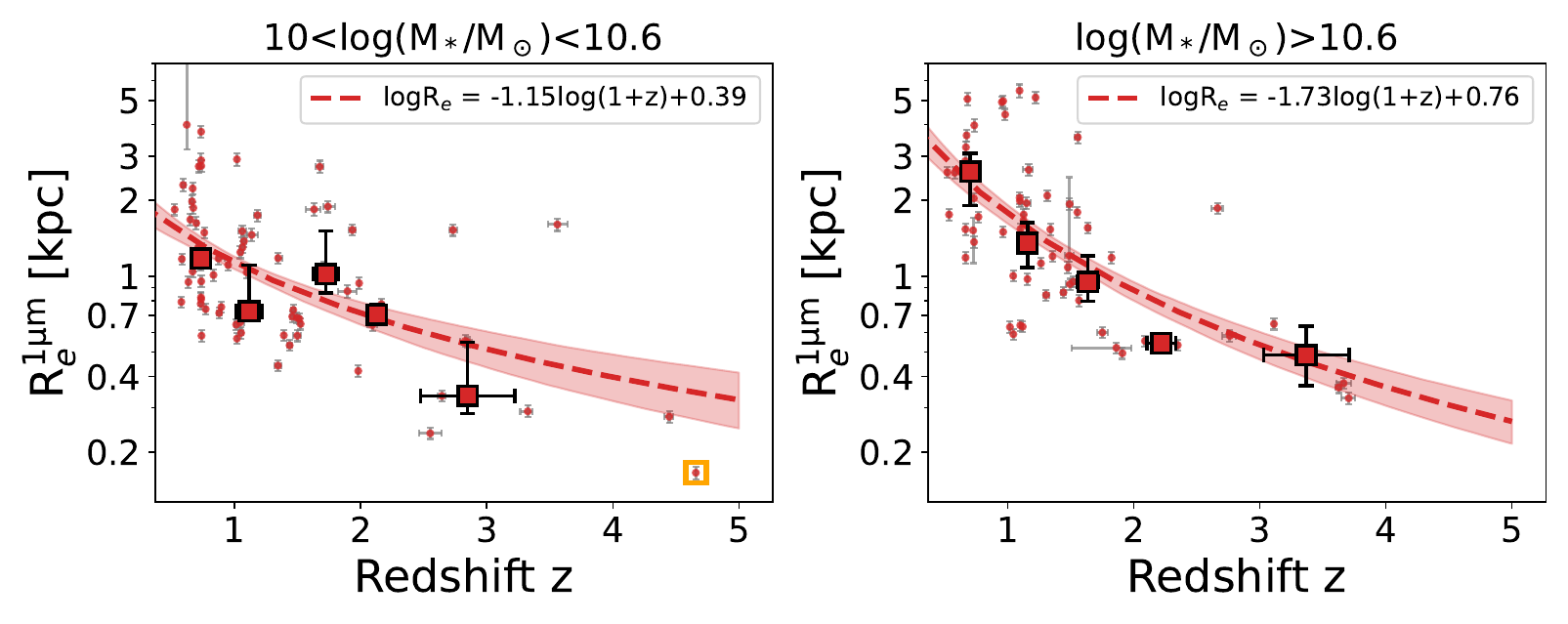}
    \caption{Stellar-mass dependence of the size evolution at rest-1$\micron$. The left panel shows the size evolution of quiescent galaxies with $\log (M_*/M_\sun)=10-10.6$, while the right shows the evolution for those with $\log (M_*/M_\sun)>10.6$. Little dots are individual quiescent galaxies, and large squares are the median size of each redshift bin (the same as Figure \ref{fig:size_evo}). The dashed red line shows the best-fit evolution (Equation \ref{equ:size}), and the shaded region marks the 1$\sigma$ uncertainty estimated using the bootstrap Monte Carlo method. The size evolution is faster for more massive quiescent galaxies. The similar is also found for the size evolution at rest-0.3$\micron$ and 0.5$\micron$ (Table \ref{tab:evo_M}). Like previous figures, GS-9209 is marked with the orange square. }
    \label{fig:size_Mdep}
\end{figure*}

We continue to divide the sample into two mass bins using the median stellar mass $\langle\log (M_*/M_\sun)\rangle=10.6$. We present the size evolution of the two mass bins in Figure \ref{fig:size_Mdep} and Table \ref{tab:evo_M}. We observe a dependence of the size evolution on stellar mass, where the evolution is faster -- more negative $\beta$ (Equation \ref{equ:size}) -- for more massive quiescent galaxies. This finding does not depend on the rest-frame wavelength at which the sizes are measured (Table \ref{tab:evo_M}). 

\begin{figure}
    \centering
    \includegraphics[width=0.47\textwidth]{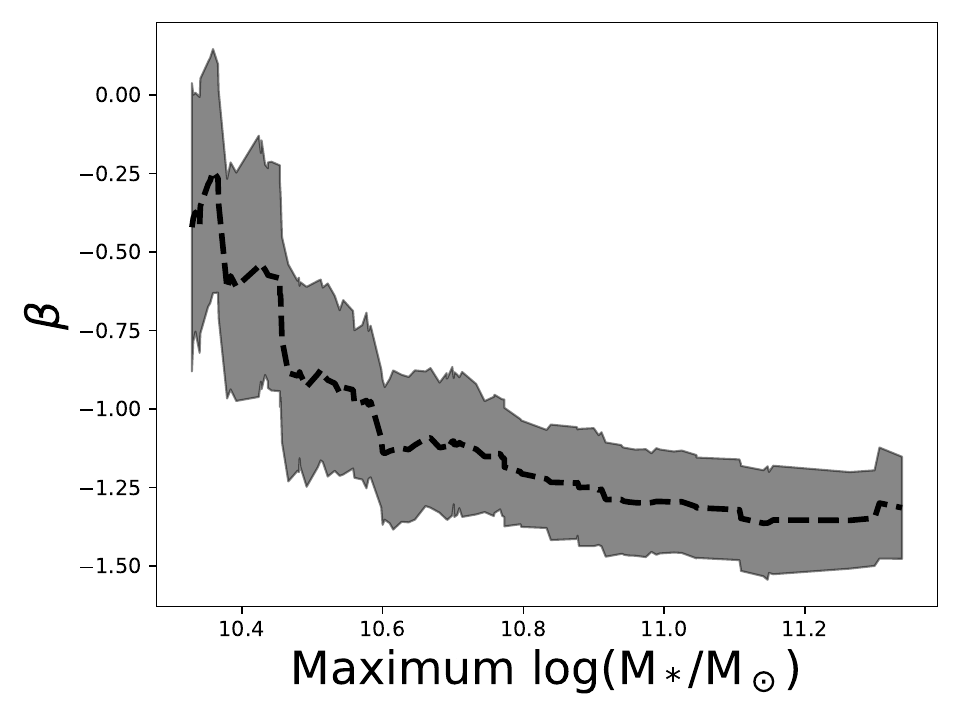}
    \caption{The change of the best-fit $\beta$ (Equation \ref{equ:size}) as more massive quiescent galaxies are included in the fit (black solid line). The x-axis shows the maximum stellar mass of the quiescent galaxies included in the fit (see Section \ref{sec:pace} for details). The gray shaded regions mark the 1$\sigma$ uncertainty estimated using the bootstrap Monte Carlo method. As more massive quiescent galaxies add to the fit, the slope $\beta$ becomes more negative, i.e. the size evolution is faster for more massive quiescent galaxies.}
    \label{fig:beta_Mmax}
\end{figure}

We further check the robustness of this stellar-mass dependence. Considering the relatively small sample size of this study (161 galaxies), instead of dividing the sample into more stellar mass bins, we  first sort stellar masses of individual quiescent galaxies into an increasing order. Then, starting from the first 30 of the sorted sample, we keep adding more massive quiescent galaxies into the fit of size evolution. We finally study the change of $\beta$ as a function of the maximum stellar mass of quiescent galaxies included in the fit. As Figure \ref{fig:beta_Mmax} shows, $\beta$ becomes increasingly negative as more massive quiescent galaxies included, confirming that the size evolution is faster for more massive quiescent galaxies. In Section \ref{dis:drivers}, we will discuss in detail the implications of this finding for the physical mechanisms governing the size evolution of massive quiescent galaxies. 

\begin{table}[]
\centering
    \caption{Stellar-mass dependence of the size evolution of massive quiescent galaxies (Equation \ref{equ:size}).}
    \begin{tabular}{| c | c c | c c |}
    \toprule
          \multirow{2}{*}{$\lambda_{\rm{rest}}$} &  \multicolumn{2}{c|}{$M_*=10^{10}\sim10^{10.6}M_\sun$} & \multicolumn{2}{c|}{$M_*>10^{10.6}M_\sun$}  \\  
          & $\beta$ & $C$ & $\beta$ & $C$ \\
     \hline
        $0.3\micron$ &  -1.06$_{-0.33}^{+0.31}$ & 0.50$_{-0.12}^{+0.12}$ &  -1.68$_{-0.28}^{+0.27}$ & 0.87$_{-0.14}^{+0.14}$ \\  
    \hline
        
        $0.5\micron$ &  -1.04$_{-0.26}^{+0.25}$ & 0.43$_{-0.09}^{+0.08}$ &  -1.60$_{-0.22}^{+0.29}$ & 0.79$_{-0.11}^{+0.09}$ \\
    \hline
        
        $1\micron$ &  -1.15$_{-0.22}^{+0.25}$ & 0.38$_{-0.07}^{+0.08}$ &  -1.73$_{-0.20}^{+0.24}$ & 0.76$_{-0.09}^{+0.09}$\\  
    \hline
    \end{tabular}
    
    \label{tab:evo_M}
\end{table}

\subsubsection{The difference in size at different wavelengths} \label{sec:interc}

The intercept $C$ (Equation \ref{equ:size}) of the size evolution depends on rest-frame wavelengths (Figure \ref{fig:size_evo}). Massive quiescent galaxies are larger at shorter wavelengths: relative to their rest-1$\micron$ sizes, on average, quiescent galaxies are larger by 45\% at rest-frame 0.3$\micron$ and 15\% at rest-frame 0.5$\micron$. Our results are in quantitative agreement with other recent JWST studies \citep{Suess2022, vanderWel2023}, showing that the stellar-mass distribution of quiescent galaxies is significantly more compact than their rest-frame UV and optical light distributions, caused by the radial changes in stellar-population properties such as stellar age, metallicity and dust attenuation \citep[e.g.,][]{Suess2022, Miller2023,vanderWel2023,Suess2023}.

\begin{figure*}
    \centering
    \includegraphics[width=0.97\textwidth]{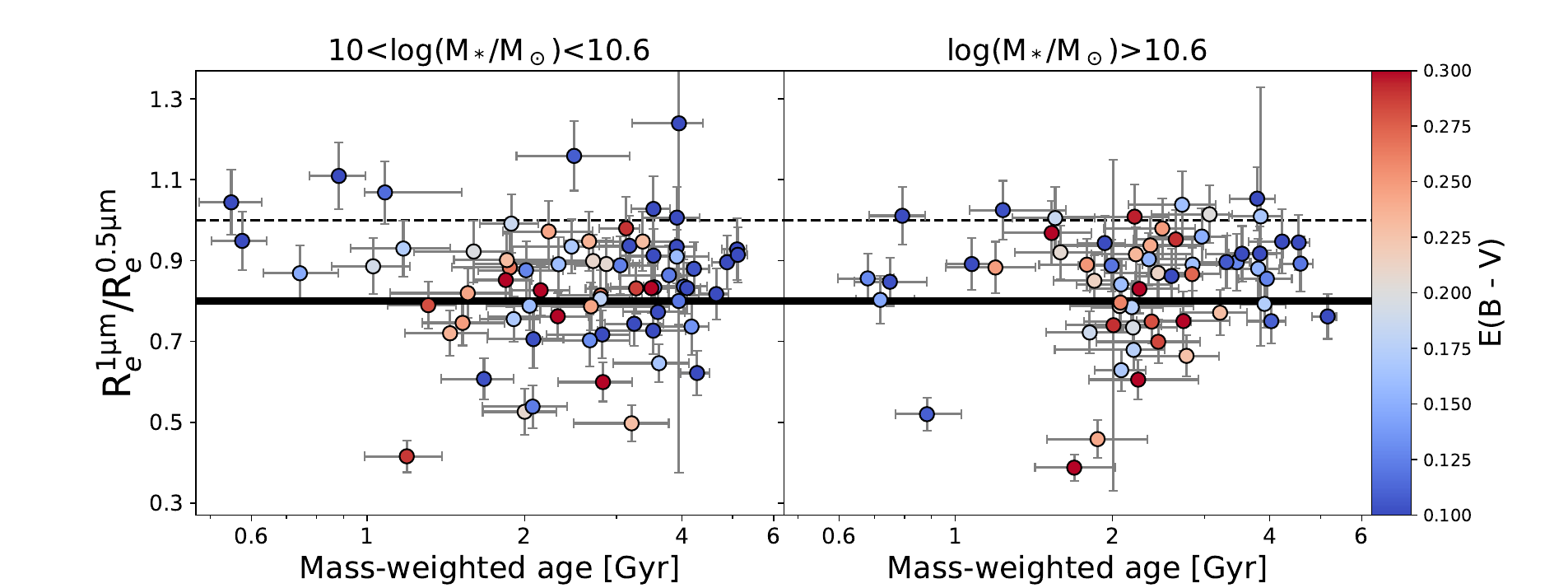}
    \includegraphics[width=0.97\textwidth]{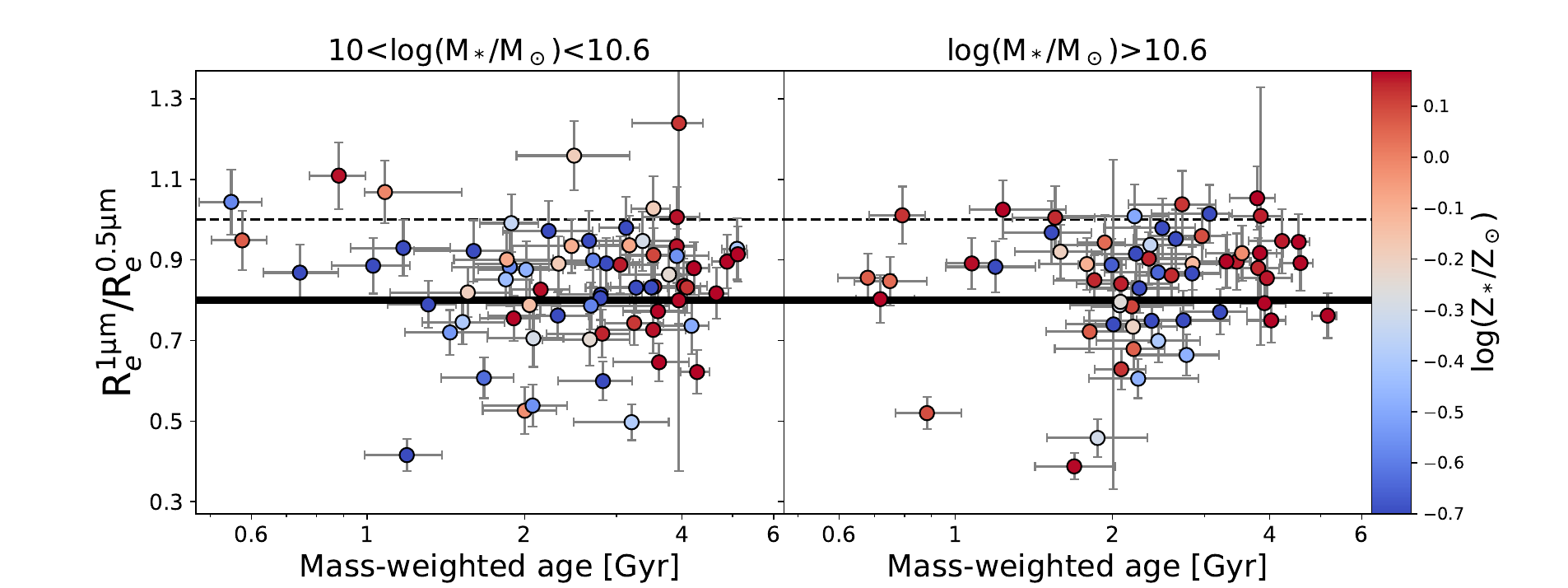}
    \caption{The ratio of rest-1$\micron$ size to rest-0.5$\micron$ size as a function of mass-weighted stellar age. In the top panel galaxies are color coded according to the color excess $\rm{E(B-V)}$, while in the bottom panel galaxies are color coded according to the stellar metallicity $Z_*$. The horizontal dashed line marks R$_e^{1\micron}$/R$_e^{0.5\micron}=1$, i.e. no evidence of color gradients. The black solid line marks R$_e^{1\micron}$/R$_e^{0.5\micron}=0.8$, about the median value of all quiescent galaxies in this study. Younger quiescent galaxies in general have rest-optical to rest-NIR size ratios closer to unity, i.e. an indication of flatter color gradients. As they become older, we observe an emergence of quiescent galaxies with large color gradients, i.e. R$_e^{1\micron}$/R$_e^{0.5\micron}<0.8$. }
    \label{fig:ratio}
\end{figure*}

We attempt to constrain the key physical quantity driving the differences in morphology at different rest-wavelengths. In Figure \ref{fig:ratio}, we plot the rest-NIR to optical size ratio,  i.e. R$_e^{1\micron}$/R$_e^{0.5\micron}$, as a function of the mass-weighted stellar age. Because the strength of color gradients depends on stellar mass \citep{Tortora2010,Suess2019,vanderWel2023}, we separate our sample into two stellar-mass bins using $\langle \log (M_*/M_\sun)\rangle=10.6$, the median value of the entire sample. Regardless of stellar mass, younger quiescent galaxies with ages $\le1.5$ Gyr have relatively flat color gradients, i.e. R$_e^{1\micron}$/R$_e^{0.5\micron}$ closer to 1, in broad agreement with studies of post-starburst galaxies \citep{Maltby2018,Suess2020}. As galaxies become older ($>1.5$ Gyr), we observe an increase in the number of quiescent galaxies with large difference in rest-optical and NIR morphologies (R$_e^{1\micron}$/R$_e^{0.5\micron}<0.8$). Color gradients of older quiescent galaxies are more negative, i.e. their centers are redder. This is consistent with the general picture of inside-out growth of massive galaxies, followed by inside-out quenching, where galaxies built up their centers earlier than outskirts \citep[e.g.][]{Tacchella2015,Nelson2016,Ji2023}. 

We see tentative evidence that old ($>$1.5 Gyr), higher-mass quiescent galaxies have flatter color gradients than lower-mass ones. The median size ratio $\langle$R$_e^{1\micron}$/R$_e^{0.5\micron}$$\rangle$ is 0.87$\pm$0.02 for $\log (M_*/M_\sun)>10.6$ quiescent galaxies, whereas it is 0.83$\pm$0.02 for $10<\log (M_*/M_\sun)<10.6$ ones  (the quoted uncertainty is estimated following the bootstrap Monte Carlo method). This is in line with the studies of nearby massive ($\log(M_*/M_\sun)>10$) galaxies from SDSS: \citet{Tortora2010} found that the most massive ($\log(M_*/M_\sun)>11$) early type galaxies have almost flat color gradients which become increasingly more negative as early type galaxies become less massive. This finding needs to be tested with future larger samples of massive quiescent galaxies at high redshifts. 

We also check the relationship of R$_e^{1\micron}$/R$_e^{0.5\micron}$ with dust attenuation $\rm{E(B-V)}$ and stellar metallicity $Z_*$. We do not see clear trends, although this may be a result of the degeneracy between $\rm{E(B-V)}$  and $Z_*$ that makes the inference of them individually highly uncertain with photometric data only. We defer detailed analysis of the spatially resolved stellar populations of the quiescent galaxies to a future study aimed at dissecting the physical mechanism(s) behind this finding. 

Finally, in addition to R$_e^{1\micron}$/R$_e^{0.5\micron}$, we have also studied the trends discussed above using R$_e^{1\micron}$/R$_e^{0.3\micron}$. We do not find any substantial changes in our conclusions, although the measurement of R$_e^{1\micron}$/R$_e^{0.3\micron}$ is noisier than R$_e^{1\micron}$/R$_e^{0.5\micron}$ as quiescent galaxies are fainter in rest-UV than in rest-optical.

\subsubsection{Comparing the size evolution of this work with other studies}

We first compare our results with HST studies. \citet{vanderWel2014} combined all five CANDELS fields to measure the rest-0.5$\micron$ size evolution over $0<z<2.5$. Our new measurements obtain $\beta\approx-1.25\pm0.20$ (Table \ref{tab:size_evo}) which is broadly consistent with \citet{vanderWel2014} who found $-1.3\lesssim\beta\lesssim-1$ for $\log (M_*/M_\sun)>10$ quiescent galaxies (see their Table 2). This shows that the size evolution of quiescent galaxies established based on rest-optical HST imaging still holds with the deeper, higher resolution JWST/NIRCam observations. 

More specifically, regarding the size evolution of quiescent galaxies with $\log (M_*/M_\sun)=10-10.6$, we obtain $\beta=-1.04^{+0.25}_{-0.26}$ (Table \ref{tab:evo_M}), which is in quantitative agreement with \citet{vanderWel2014} who found $\beta=-1.01\pm0.06$ for quiescent galaxies of similar stellar masses. Regarding the higher-mass ones, i.e. $\log (M_*/M_\sun)>10.6$, we obtain $\beta=-1.60^{+0.29}_{-0.22}$, compared to $\beta=-1.32\pm0.21$ from \citet{vanderWel2014}. Statistically, this difference in $\beta$ is marginal, $\lesssim1\sigma$. Nonetheless, we still investigate what causes the marginally steeper size evolution we found for the $\log (M_*/M_\sun)>10.6$ quiescent galaxies. First, this is not due to the systematics in the size measures between JWST and HST, as we found consistent sizes between the measurements using NIRCam/F150W and HST/F160W imaging (Section \ref{sec:galfit}). Second, we note that with HST \citet{vanderWel2014} was only able to measure the rest-optical size evolution up to $z=2.5$, whereas with JADES we can now extend the size measurements of quiescent galaxies beyond $z>2.5$. If we only include $z<2.5$ galaxies in the fit, we will still find a similarly steeper size evolution than \citet{vanderWel2014}. Therefore the steeper size evolution from our measurements is not due to including $z>2.5$ quiescent galaxies either. The main cause for the steeper size evolution seems to be sample selection. The analysis of \citet{vanderWel2014} included all UVJ quiescent galaxies selected based on 3D-HST rest-frame colors, while we removed from our analysis 36 out of 197 galaxies which do not satisfy the UVJ criteria any more after we re-fit their SEDs with new photometry from JADES (Section \ref{sec:sample}). If those 36 galaxies were not removed, we would have found $\beta=-1.30\pm0.27$ which is fully consistent with \citet{vanderWel2014}. We note that the majority -- 22 out of 36 -- of the removed galaxies are at $z>1.5$, most of which are near the UVJ selection boundaries. Interestingly, the removed galaxies generally have more extended morphologies than those retained in our final sample, indicating a strong correlation between the size and star formation properties of galaxies at Cosmic Noon \citep[][just to name a few]{Wuyts2011,Williams2010,Patel2013,vanderWel2014,Ji2023}.

We now compare our results with a very recent JWST study done by \citet{vanderWel2023} who measured the size evolution of massive quiescent galaxies using NIRCam data from the CEERS survey. \citet{vanderWel2023} obtained $\beta=-1.70\pm0.10$ for $\log (M_*/M_\sun)>11$, $0.5<z<2.3$ quiescent galaxies. If we fit the size evolution only for a subsample of the same redshift and stellar-mass ranges, we will obtain a fully consistent $\beta\approx-1.75$. Moreover, \citet{vanderWel2023} found that the pace of the size evolution does not depend on size proxies used (e.g., half-light vs half-mass). This is also similar to our finding above that $\beta$ does not depend on rest-frame wavelengths where the sizes are measured (Table \ref{tab:size_evo} and Figure \ref{fig:size_evo}). 

\begin{figure}
    \centering
    \includegraphics[width=0.47\textwidth]{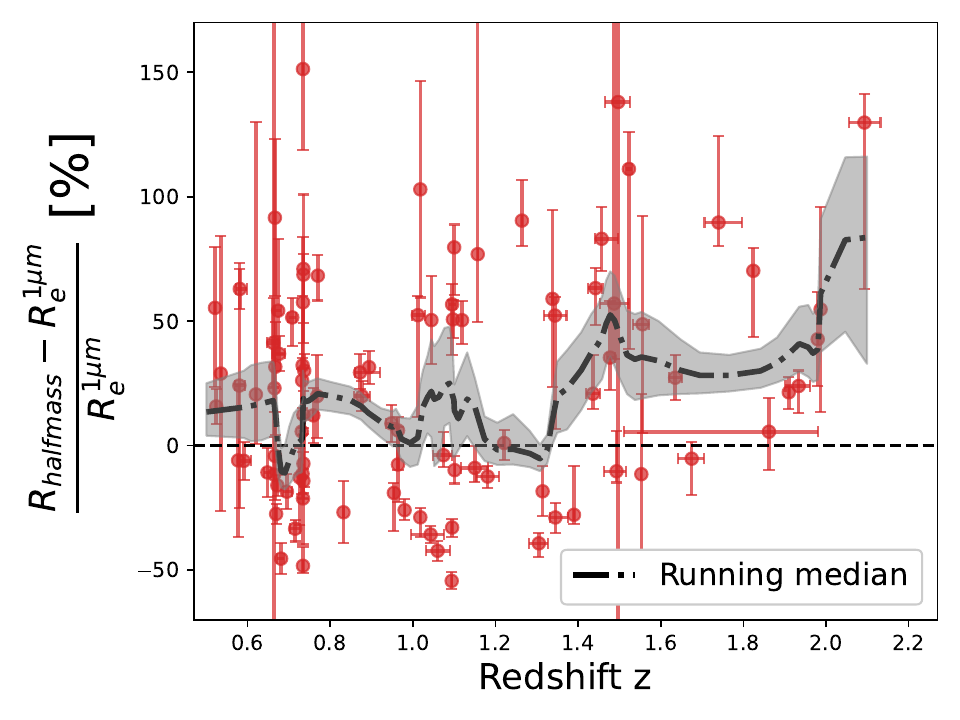}
    \caption{Relative difference between $R_e^{1\mu m}$ and the half-mass sizes from \citet{Suess2019} as a function of redshift. The grey dashdotted line and shaded region show the running median the its 1$\sigma$ uncertainty. The uncertainty of the relative difference (y-axis) is dominated by the uncertainty of half-mass size measurements from \citet{Suess2019}.}
    \label{fig:deltar_z}
\end{figure}

Our results -- and hence those of \citealt{vanderWel2023} -- then suggest significant size evolution of massive quiescent galaxies beyond $z>1$, which is in tension with some previous studies based on HST imaging: \citet{Suess2020} and \citet{Miller2023} found that the half-mass size evolution of quiescent galaxies is slower than their rest-optical, half-light size evolution, suggesting that the latter is mostly due to stellar mass-to-light radio gradients, rather than the intrinsic growth of galaxies' sizes. In Figure \ref{fig:deltar_z},  for the subsample having $R_{\rm{halfmass}}$ from \citet{Suess2020}, we plot the relative difference between $R_{\rm{halfmass}}$ and $R_e^{1\mu m}$ as a function of redshift. We also plot the running median and its uncertainty estimated following the bootstrap Monte Carlo method described earlier (Section \ref{sec:size_evo}). Up to $z\sim1.2$, the median of the relative difference remains roughly flat, with $R_{\rm{halfmass}}$ being larger than $R_e^{1\mu m}$ by $14\pm8\%$. Although the sample size of quiescent galaxies having both $R_{\rm{halfmass}}$ and $R_e^{1\mu m}$ measurements is limiting at $z>1.2$, we see evidence that $R_{\rm{halfmass}}$ becomes increasingly larger than $R_e^{1\mu m}$ towards higher redshifts. This explains the shallower half-mass size evolution at $z>1$ reported by those earlier HST studies. \citet[][see their Appendix A]{vanderWel2023} also attempted to address this tension regarding the half-mass size evolution at $z>1$ by comparing the light-to-mass weighed size ratios from their measurements with those from \citet{Suess2020}. Similar to what we found in Figure \ref{fig:deltar_z}, \citealt{vanderWel2023} showed that, while the light-to-mass weighed size ratio from \citet{Suess2020} becomes increasingly larger at $z>1.5$, they did not see this trend in their new measurements using CEERS data. 

We point out that the methods used to derive the half-mass sizes are dramatically different in the literature, which can lead to inconsistent conclusions about the size evolution of quiescent galaxies. For example, also using HST imaging, \citet{Mosleh2017} adopted a different approach to account for the radial gradient of the stellar mass-to-light ratio. Unlike \citet{Suess2020} and \citet{Miller2023}, \citet{Mosleh2017} found that half-mass sizes of quiescent galaxies significantly evolve at $z>1$. In the future, dedicated analysis of different methodologies in deriving the half-mass sizes, and larger samples of quiescent galaxies at $z>2$ are needed to finally address this tension between JWST and HST studies. 

\subsection{Stellar mass-size Relationship of Massive Quiescent Galaxies}\label{sec:mass-size}

We now present the mass-size relationship of massive quiescent galaxies. To characterize it, we assume the functional form of 
\begin{equation}
R_e^{\rm{1\mu m}}\, [\rm{kpc}] = A\cdot \left(\frac{M_*}{10^{10.7}M_\sun}\right)^{\Gamma},
\label{equ:mass-size}
\end{equation}
where both the amplitude $A$ and power law index $\Gamma$ depend on redshifts following $A = A_0(1+z)^\alpha$ and $\Gamma = \Gamma_0(1+z)^\gamma$. It has been shown by previous studies that the mass-size relationship of quiescent galaxies cannot be well described by a single power law where the relationship becomes flattened below the pivot mass $\sim 10^{10.4}M_\sun$ \citep[e.g.][]{vanderWel2014,Cutler2022,vanderWel2023}. Therefore, we only fit Equation \ref{equ:mass-size} for galaxies with $M_*>10^{10.4}M_\sun$. Like before, we estimate the uncertainties of the fitted parameters using the bootstrap Monte Carlo method: We bootstrap the entire sample, use normal distributions to Monte Carlo resample the $M_*$ and $R_e^{\rm{1\mu m}}$ measures using their corresponding uncertainties, and finally fit the mass-size relationship. We repeat this procedure 10000 times and use the range between 16th and 84th percentiles as $1\sigma$ uncertainties.

\begin{figure*}
    \centering
    \includegraphics[width=0.997\textwidth]{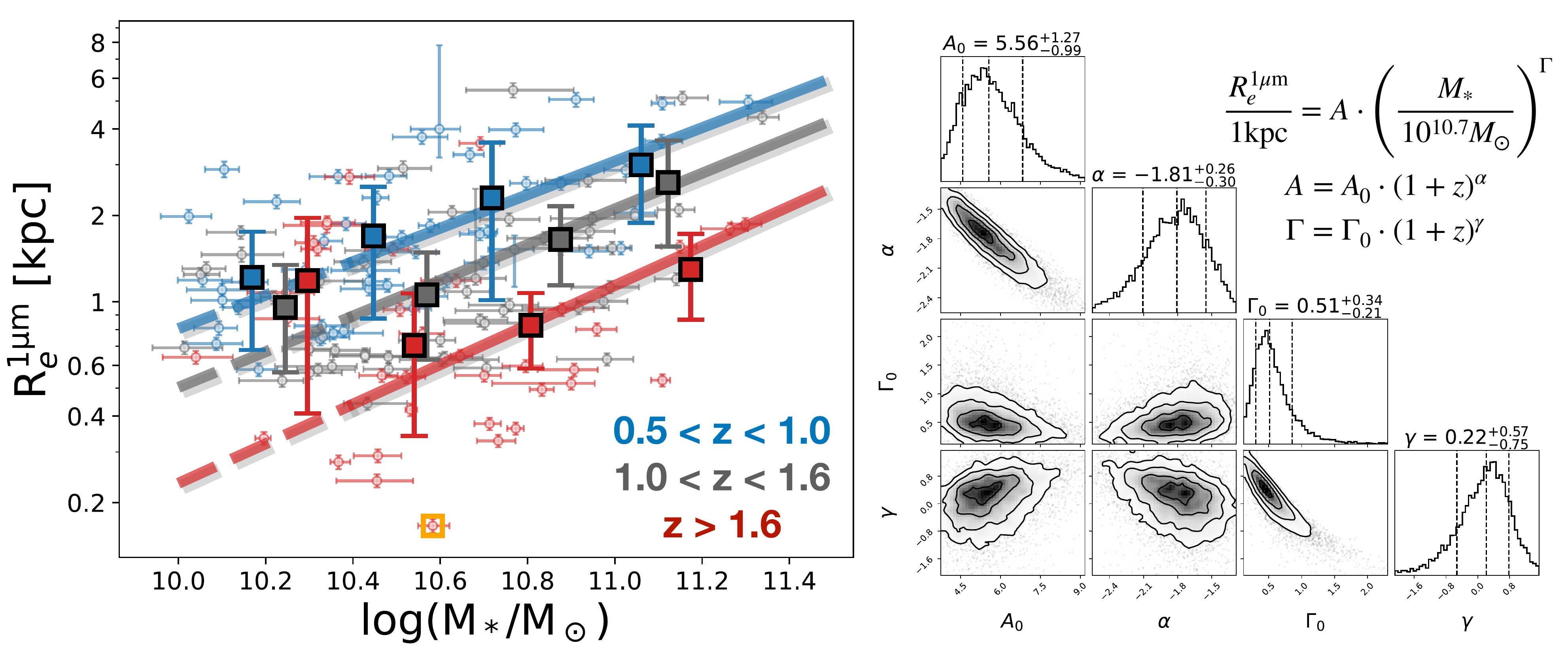}
    \caption{Stellar mass-size relationship of massive quiescent galaxies. In the left panel we show the scatter plot of stellar masses vs rest-1$\micron$ sizes, where the entire sample is divided into three groups using redshift terciles. Small dots show the individual measurements, and large squares show the median sizes of individual stellar-mass bins. For galaxies with $\log (M_*/M_\sun)\ge10.4$, we fit the redshift-dependent mass-size relationship using the functional form shown on the right (see Section \ref{sec:mass-size} for details). Solid lines show the best-fit relationships at the median redshifts of the three groups. We also extrapolate the relation to $\log (M_*/M_\sun)<10.4$, shown as dashed lines.  The right panel shows the corner plots for the fitted parameters of the mass-size relationship. Like before, GS-9209 is marked with the orange square.}
    \label{fig:mass_size}
\end{figure*}

In Figure \ref{fig:mass_size} we show the mass-size relationship. The amplitude $A$ strongly evolves with redshift following $(1+z)^{-1.8}$, showing that at a fixed stellar mass, quiescent galaxies are significantly larger at lower redshifts. The slope of the mass-size relationship is $\Gamma_0\sim0.5$ and it does not strongly depend on redshift with $\gamma$ being consistent with 0, in broad agreement with previous studies \citep{vanderWel2014,Mowla2019}. We note, however, that the current sample size of $z>3$ quiescent galaxies is very limiting and future larger samples are needed to check if the slope of the mass-size relationship persists towards higher redshifts. Finally, for completeness we also report the best-fit mass-size relationships of the rest-frame 0.3$\micron$ and 0.5$\micron$ sizes in Table \ref{tab:mass-size}. 

\begin{table}[]
\centering
    \caption{The best-fit mass-size relationship of massive quiescent galaxies (Equation \ref{equ:mass-size}).}
    \begin{tabular}{| c | c c c c |}
    \toprule
    $\lambda_{\rm{rest}}$ & $A_0$ [kpc] & $\alpha$ & $\Gamma_0$ & $\gamma$ \\
    \hline
        $0.3\micron$ &  8.43$_{-1.58}^{+2.10}$ & -1.78$_{-0.30}^{+0.26}$ &  0.49$_{-0.28}^{+0.49}$ & 0.17$_{-0.94}^{+0.78}$ \\  
    \hline
        
        $0.5\micron$ &  5.04$_{-0.73}^{+1.02}$ & -1.41$_{-0.26}^{+0.22}$ &  0.48$_{-0.20}^{+0.31}$ & 0.33$_{-0.71}^{+0.63}$ \\
    \hline
        
        $1\micron$ &  5.56$_{-0.99}^{+1.27}$ & -1.81$_{-0.30}^{+0.26}$ &  0.51$_{-0.21}^{+0.34}$ & 0.22$_{-0.75}^{+0.57}$\\  
    \hline
    \end{tabular}
    
    \label{tab:mass-size}
\end{table}

\section{Discussion}

With the deep, multi-band JWST/NIRCam imaging data from JADES, we have studied the size evolution of a robust sample of massive quiescent galaxies with $\log (M_*/M_\sun) > 10$ over $0.5 < z < 5$. Regardless of the rest-frame wavelengths where the sizes are measured, we observed strong size evolution of massive quiescent galaxies. This evolution depends on stellar mass, with more massive quiescent galaxies having faster evolution in size. In the following, we will discuss in detail the implications of the observed size evolution on the formation and evolution of quiescent galaxies across cosmic time. We will also discuss our findings with regard to the quiescent galaxies at $z>3$.

\subsection{Physical Mechanisms Driving the Size Evolution of Quiescent Galaxies} \label{dis:drivers}

Two broad categories of physical mechanisms can drive the strong size evolution of quiescent galaxies without significantly increasing their stellar masses. First, in $\Lambda$CDM cosmology, at a fixed mass galaxies formed later are expected to be larger, the progenitor effect \citep{Carollo2013,Lilly2016,Ji2022}. It is therefore possible that the apparent size evolution of quiescent galaxies is driven by the addition of newly quenched, larger galaxies at lower redshifts. Alternatively, post-quenching growth --  particularly mechanisms such as gas-poor minor mergers that can efficiently grow sizes without significantly increasing masses -- can also explain the observed strong size evolution \citep[e.g.][]{Bezanson2009, Naab2009}. 

Regarding the progenitor effect, there is a simple prediction for the pace of apparent size evolution. Specifically, by construction, the Virial radius ($R_{\rm{vir}}$) of a dark matter halo is set by the threshold of density contrast -- the ratio of density to the mean density of the Universe -- at the redshift of halo collapse ($z_{\rm{form}}^{\rm{h}}$).  Because the density contrast threshold has a very weak time dependence, we then expect that, at a fixed halo mass, $R_{\rm{vir}}$ should linearly scale with the size of the Universe, i.e. $R_{\rm{vir}}\propto H(z_{\rm{form}}^{\rm{h}})^{-2/3}$ where $H(z_{\rm{form}}^{\rm{h}})$ is the Hubble constant at $z_{\rm{form}}^{\rm{h}}$ \citep{Mo2010}. Given that $H(z)\sim(1+z)^{1.4}$ at Cosmic Noon $z\sim2$, we would then expect $R_{\rm{Vir}}\propto (1+z_{\rm{form}}^{\rm{h}})^{-1}$. Finally, because the ratio of the galaxy $R_e$ to the halo $R_{\rm{Vir}}$ is roughly constant, as supported by both theories \citep[e.g.,][]{Mo2010} and observations \citep[e.g.,][]{Kravtsov2013,Somerville2018}, the prediction for the apparent size evolution driven by the progenitor effect is then $R_e \propto (1+z_{\rm{form}}^{\rm{h}})^{-1}$. 

\subsubsection{Lower-mass quiescent galaxies with $10<\log (M_*/M_\sun)<10.6$} \label{sec:dis_size_evo_lowM}

The size evolution of $10^{10}M_\sun<M_*<10^{10.6}M_\sun$ quiescent galaxies follows $(1+z)^{-1.15_{-0.22}^{+0.25}}$ (Section \ref{sec:size_evo}), which is broadly consistent with $R_e\propto(1+z)^{-1}$ and hence is in line with the progenitor effect. 

Before moving forward, we clarify several things with regard to the comparison between the observed size evolution and the prediction of the progenitor effect. First, the prediction is made for galaxies at a fixed halo mass, rather than a fixed stellar mass. However, we note that the stellar-mass range considered here -- $10^{10}M_\sun<M_*<10^{10.6}M_\sun$ -- is where galaxies' halo masses do not change substantially over $0.5<z<5$ \citep[see e.g., the left panel of Figure 7 in ][]{Behroozi2013}. If we use the relationship between stellar mass and halo mass from \citet{Behroozi2013,Behroozi2013b}\footnote{At $z\le5$, the data release of \citet{Behroozi2013,Behroozi2013b} only contains six redshift slices, i.e. $z=0.1, 1, 2, 3, 4,5$. We therefore interpolate those slices to derive the halo mass of a galaxy at its exact redshift value.}, we find that the halo masses of our quiescent samples at different redshifts are very similar, with the median halo mass being $10^{12}M_\sun$ for the $z>2$ subsample compared to $10^{11.9}M_\sun$ for the $z<2$ subsample.  

\begin{figure}
    \centering
    \includegraphics[width=0.47\textwidth]{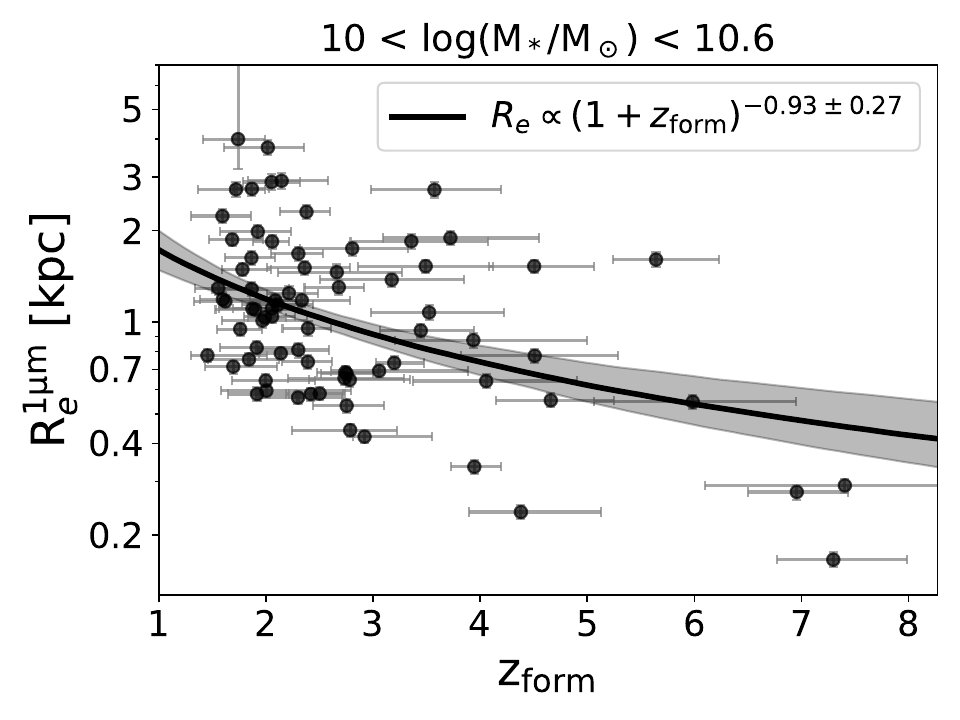}
    \caption{$R_e^{1\mu m}$ as a function of formation redshift ($z_{\rm{form}}$) for quiescent galaxies with $10<\log (M_*/M_\sun)<10.6$. The relationship between $R_e^{1\mu m}$ and $z_{\rm{form}}$ is broadly consistent with $(1+z_{\rm{form}})^{-1}$, in line with the expectation from the progenitor effect (Section \ref{sec:dis_size_evo_lowM}). }
    \label{fig:size_zf}
\end{figure}

Second, the halo formation theory detailed above, if applicable, suggests that the size of galaxies should more tightly correlate with $z_{\rm{form}}^{\rm{h}}$, the time of halo formation, than with $z$, the time of observation. Following \citet{Ji2022}, instead of using $z$, we study the relationship between $R_e$ and the formation redshift, $z_{\rm{form}}$, which is defined as the redshift when a galaxy reached its mass-weighted age. As Figure \ref{fig:size_zf} shows, the best-fit relationship is $R_e^{1\mu m} \propto (1+z_{\rm{form}})^{-0.93\pm0.27}$. Compared to $\beta=-1.15$ obtained for the relationship between $R_e^{1\mu m}$ and $z$ (Section \ref{sec:size_evo}), using $z_{\rm{form}}$ makes the slope of the size evolution closer to $-1$, i.e. the prediction of the progenitor effect. We note, however, that using $z_{\rm{form}}$ as a proxy for $z_{\rm{form}}^{\rm{h}}$ is purely from an empirical standpoint, and very likely there are better metrics which however is beyond the scope of this work.

Finally, in addition to fitting the size evolution as a function of $(1+z)$, which is widely used in the literature but arguably not the best physically motivated form in terms of halo formation (see description above and also \citealt{vanderWel2014}), we also fit the size evolution as a function of the Hubble constant $H(z)$, i.e., $R_e\propto H(z)^{\beta_H}$. We find $\beta_H =-0.87_{-0.15}^{+0.21}$ for the relationship between $R_e^{1\mu m}$ and $H(z)$, and $\beta_H =-0.64_{-0.17}^{+0.23}$ for the relationship between $R_e^{1\mu m}$ and $H(z_{\rm{form}})$. Similarly, the best-fit $\beta_H$ of the relationship between $R_e^{1\mu m}$ and $z_{\rm{form}}$ is closer to $\beta_H = -2/3$ predicted by the progenitor effect.

Obviously, the interpretation of the observed slope of size evolution is highly degenerated, as the slope might reflect either a single, dominated physical mechanism or the interplay among several mechanisms that can mimic a similar size evolution. Nonetheless, particularly because the slope becomes closer to the predicted value of the progenitor effect after switching from $z$ to $z_{\rm{form}}$, we believe that the progenitor effect may be the primary mechanism, if not the only one, driving the apparent size evolution observed in the quiescent galaxies with $10<\log (M_*/M_\sun)<10.6$. If this indeed is not a coincidence, our findings will then indicate that the size of quiescent galaxies evolves at a similar pace following their dark matter halos. Because of this, we  speculate that dark matter halos play a key role, if not the only one, in the quenching and structural evolution of galaxies of this stellar-mass range, which has been suggested by some theoretical works \citep[e.g.,][]{Birnboim2003,Cattaneo2006,Dekel2006,Chen2020}. 

\subsubsection{Higher-mass quiescent galaxies with $\log (M_*/M_\sun)>10.6$} \label{sec:dis_size_evo_highM}

Relative to the lower-mass ones, the size of higher-mass  quiescent galaxies increases faster with cosmic time (Section \ref{sec:pace}). The higher-mass quiescent galaxies also have a steeper relationship between $R_e^{1\mu m}$ and $z_{\rm{form}}$ with a slope of $-1.57\pm0.26$. These deviate from the prediction of the progenitor effect alone, suggesting additional mechanism(s) contributing to the structural growth. 

Higher-mass quiescent galaxies are hosted by more massive halos formed in larger overdensities \citep{Mo2010}. Consequently, they are expected to experience more merger events than lower-mass galaxies \citep[e.g.,][]{Peng2010,Hopkins2010,RodriguezGomez2015,OLeary2021}. Therefore, a likely explanation for their apparent steeper size evolution is the increasing contribution of post-quenching size growth from mergers, especially minor mergers that can grow galaxies' sizes more efficiently than major mergers for a fixed amount of mass to be added to the central galaxy \citep[e.g.,][]{Naab2009,Bezanson2009}. 

If post-quenching mergers do play a significant role in growing the size of higher-mass quiescent galaxies, one would then expect that the slope of the relationship between $R_e^{1\mu m}$ and $z_{\rm{form}}$ depends on galaxy ages, as older galaxies have more time to alter their structure through mergers. Motivated by this, we divide the higher-mass quiescent sample into the young and old subsamples. Because galaxies at high redshifts are generally younger, instead of making the division using the median age of the entire sample, we first estimate the running median of stellar ages as a function of redshift. The old (young) subsample contains galaxies older (younger) than the median stellar age of their redshifts.

\begin{figure*}
    \centering
    \includegraphics[width=0.97\textwidth]{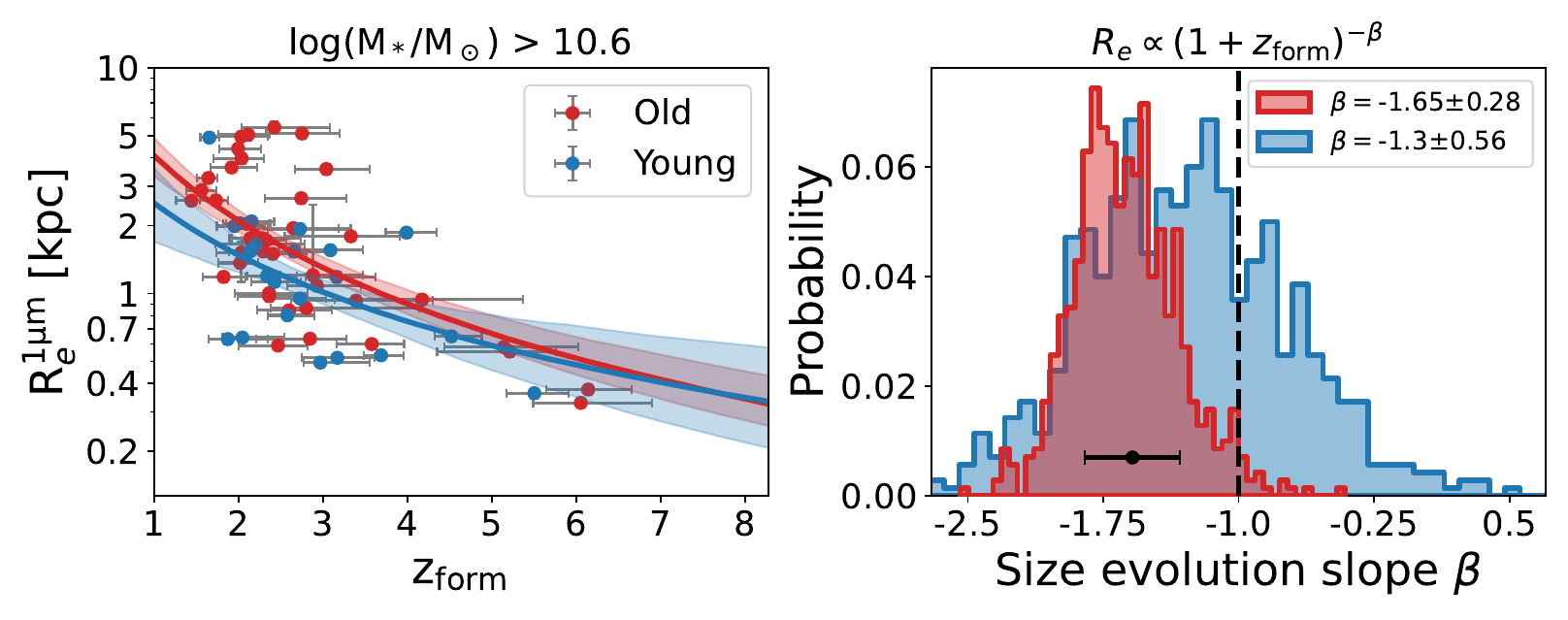}
    \caption{$R_e^{1\mu m}$ as a function of formation redshift ($z_{\rm{form}}$) for quiescent galaxies with $\log (M_*/M_\sun)>10.6$. We divide the sample into old (red) and young (blue) subsamples (Section \ref{sec:dis_size_evo_highM}). The left panel shows the best-fit relationship between $R_e^{1\mu m}$ and $z_{\rm{form}}$. The right panel shows the distributions of $\beta$ for the two subsamples from our bootstrap Monte Carlo method (Section \ref{sec:dis_size_evo_highM}). The black circle with error bars shows the median and standard deviation of $\beta$ derived for the entire $\log (M_*/M_\sun)>10.6$ sample. The vertical, black dashed line marks $\beta=-1$, the expectation from the progenitor effect. Old quiescent galaxies have a steeper $R_e^{1\mu m}$-$z_{\rm{form}}$ relationship, while the relationship of young quiescent galaxies is closer to $\beta=-1$, i.e. the progenitor effect. }
    \label{fig:size_zf_highm}
\end{figure*}

We study the relationship between $R_e^{1\mu m}$ and $z_{\rm{form}}$ for the young and old subsamples, respectively. To estimate uncertainties of the fitted parameters, we use the bootstrap Monte Carlo method that has been described previously (e.g., Section \ref{sec:size_evo} and \ref{sec:mass-size}).  As Figure \ref{fig:size_zf_highm} shows, although the current sample size is still limiting, we see evidence that the relationship between $R_e^{1\mu m}$ and $z_{\rm{form}}$ depends on stellar ages. The old subsample has a steeper (more negative) slope than the young subsample. Interestingly, although the slope ($-1.30\pm0.56$) observed in the young quiescent subsample is still somewhat steeper, it is broadly consistent with $-1$ within uncertainty, and is closer to the expectation from the progenitor effect than the old subsample. This finding is consistent with the picture that, apart from the progenitor effect, mergers and/or post-quenching continuous gas accretion drive additional, apparent size growth in very massive quiescent galaxies. As they become older, massive quiescent galaxies experience more merger events that can effectively grow their sizes by adding mass to outskirts. 

The conclusions above based on size evolution are in line with other analysis of quiescent galaxies at Cosmic Noon. First, our conclusions are consistent with the stacking analysis of quiescent galaxies' surface brightness profiles at $z\sim2$ \citep[see Section 3.4 of][]{Ji2022} and at $z\sim0$ \citep{Huang2013a,Huang2013b}, and also in broad agreement with other earlier studies, both observationally and theoretically, where they concluded the mass and size growths of massive quiescent galaxies have a significant contribution from minor mergers at $z<2$ \citep[e.g.,][]{Bezanson2009,vanDokkum2010,Oser2012,RodriguezGomez2016}. Our conclusions are further supported by a recent JWST study by \citet{Suess2023} who reported the detection of a large population of low-mass companions within 35 kpc of quiescent galaxies with $\log (M_*/M_\sun)>10.5$ at $0.5<z<3$. Our conclusions are also in line with recent ALMA observations for a small number of massive quiescent galaxies at $z\sim0.7$, where the larger-size extended ones simultaneously show evidence for minor merging both in their SFHs and newly accreted gas \citep{Spilker2018,Woodrum2022}. 

While so far we lack the kinematics constraints for our high-redshift quiescent galaxies, it is worth pointing out that the observed stellar-mass dependence of the size evolution is consistent with the dynamical studies of Early Type Galaxies (ETGs) at lower redshifts. At intermediate redshifts, \citet[][]{Bezanson2018} showed that ETGs at $z\sim0.8$ become increasingly velocity dispersion dominated as they become more massive. In addition, by studying the excess kurtosis $h_4$ of the stellar velocity distribution in $z<1$ quiescent galaxies, \citet{DEugenio2023} showed that $h_4$ -- a sensitive probe of the radial anisotropy of the stellar velocity distribution -- increases with cosmic time. At $z\sim0$, \citet[][and references therein]{Cappellari2016} showed that the distribution of the dynamical state of ETGs is bimodal -- fast and slow rotators characterized by very different ratios of rotational velocity to velocity dispersion. The two populations can be very well separated using the stellar mass, with the slow rotators being more massive than the faster rotators \citep[see a recent review by][]{Cappellari2016}. All these dynamical studies suggest that very massive galaxies after quenching continue growing their sizes via merging with other galaxies through dynamical friction. Because multiple incoherent merging events are expected to cause the loss of angular momentum \citep[e.g.,][]{Emsellem2011}, very massive galaxies eventually become verlocity dispersion dominated, i.e. slower rotators .

\subsection{The Structure of Quiescent Galaxies at $z>3$} \label{diss:zgt3}

The deep JWST/NIRCam imaging from JADES enables us to select robust samples of quiescent galaxies, and to measure their rest-optical and NIR morphologies with high precision beyond $z>3$. Despite that the current sample size is very limiting, our results suggest that the $z>3$ quiescent galaxies in our sample have very small sizes, potentially being even smaller than extrapolating the size evolution of lower-redshift quiescent galaxies (Section \ref{sec:size_evo}). A similar finding is reported by \citet{Wright2023} who found evidence that $z>3$ massive quiescent galaxies in CEERS also seem to be more compact than previously anticipated. In what follows, we take a close look at the structural properties of the $z>3$ quiescent galaxies in this study.

\subsubsection{Large stellar mass surface densities of the freshly quenched $z>3$ galaxies} \label{sec:dis_Se}

Among eleven photometrically robust quiescent galaxies at $z>3$ in our sample, eight have reliable $R_e$ estimates from  single S\'{e}rsic fitting (Section \ref{sec:galfit}). We calculate their stellar mass surface densities within the half-light radius following $\Sigma_e = M_*/(2\pi R_e^2)$. 

The remaining three galaxies have best-fit $n=8$, and they have very compact cores, as PSF spikes are clearly seen in their F444W images. While the galaxies with $n=8$ were removed from all analysis presented above, here we attempt to give reasonable $\Sigma_e$ estimates for these three galaxies as well. Instead of using $R_e$ from single S\'{e}rsic fitting, we thus use the sizes of the three galaxies from the Richardson–Lucy deconvolution approach (Section \ref{sec:test_re_zgt3}).

\begin{figure*}
    \centering
    \includegraphics[width=0.77\textwidth]{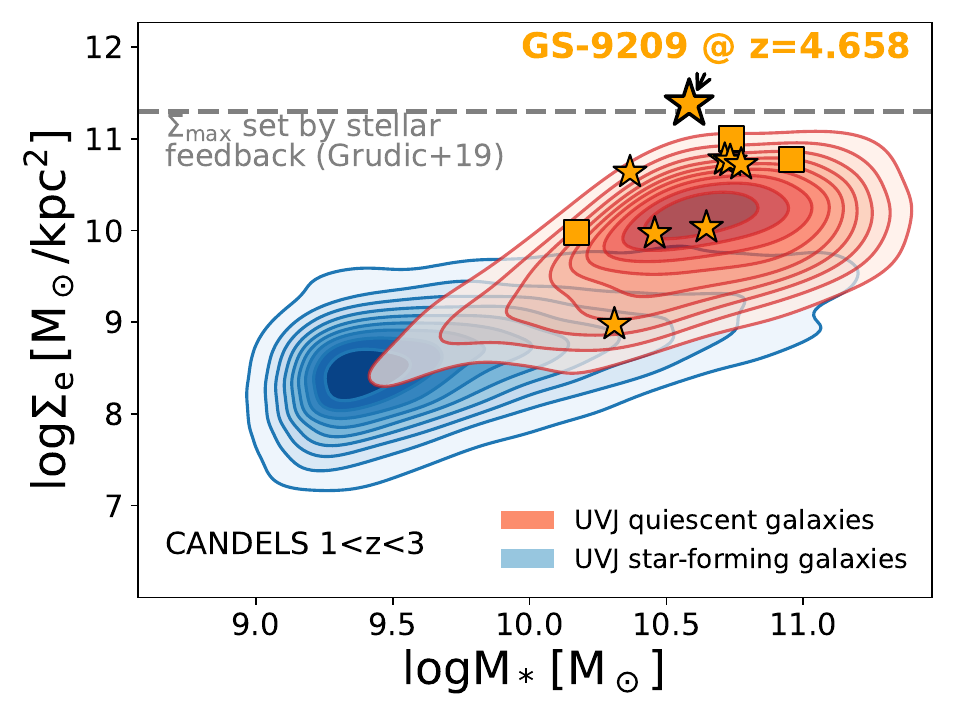}
    \caption{Stellar mass surface density within effective radius ($\Sigma_e$) vs $M_*$. Background contours show density distributions of UVJ-selected star-forming (blue) and quiescent (red) galaxies at $1<z<3$ from CANDELS. There are eleven robust quiescent galaxies at $z>3$ in our sample, among which eight (shown as orange stars) have reliable size estimates from single S\'{e}rsic fitting, including the galaxy GS-9209 that is highlighted with an enlarged symbol. For the remaining three  galaxies (shown as orange squares), we calculate their $\Sigma_e$ using alternative size measures (see Section \ref{sec:dis_Se} for details). The grey horizontal dashed line marks the maximum value of stellar mass surface density theoretically derived by \citet{Grudic2019}. }
    \label{fig:Se_mass}
\end{figure*}

We compare the $\Sigma_e$ of the $z>3$ quiescent galaxies with galaxies at $1<z<3$ from CANDELS. For CANDELS galaxies, we make use of the stellar-mass measures from \citet{Lee2018} and rest-optical sizes measured with HST/WFC3 imaging. For consistency, we thus use rest-0.5$\micron$ sizes to calculate $\Sigma_e$ for the $z>3$ galaxies, although we have already checked that any of our conclusions will not change if we use rest-1$\micron$ sizes instead. As Figure \ref{fig:Se_mass} shows, the vast majority of the $z>3$ quiescent galaxies are very compact, with $\Sigma_e > 10^{10} \,M_\sun/\rm{kpc}^2$. Such a stellar mass surface density is comparable to that observed in nuclear star clusters in nearby galaxies \citep{Norris2014}, and it is also above the averaged value found in quiescent galaxies at lower redshifts and significantly exceeds that observed in star-forming galaxies of similar stellar masses at Cosmic Noon \citep[e.g.][]{Barro2017,Ji2023}. 

We note that the presence of AGN can make galaxies appear smaller than they actually are, because the nonstellar, nucleated light from AGN can lead to a smaller size from single S\'{e}rsic fitting than the intrinsic one of stellar distribution \citep{Ji2022AGN}. Despite that AGN-hosting galaxies have already been removed from our sample through various AGN selection methods (Section \ref{sec:sample}), there could still be some missed ones because the AGN are too faint to be identified by those selection methods. One example is GS-9209 -- it shows broad H$\alpha$ emission line but is not identified as an AGN based on SED fitting -- that we will discuss in detail in Section \ref{dis:9209}. Although we currently do not know how many of the $z>3$ quiescent galaxies in our sample might host a similarly faint broad-line AGN like GS-9209, we argue this does not seem to be able to explain the strong size evolution, hence the very large $\Sigma_e$ of the $z>3$ quiescent galaxies for the following two main reasons. First, the SED shapes of the quiescent galaxies all seem to be typical for quiescent/post-starburst galaxies, with no clear evidence of hosting strong AGN of known types. Second, the broad emission lines from AGN, if present, likely can only affect the rest-0.5$\micron$ sizes (due to e.g., [\ion{O}{3}]5007\AA), while they should have much less effects on the rest-1$\micron$ sizes because this wavelength range is mostly free of strong emission lines. Yet, we still observe the very strong size evolution at rest-frame 1$\micron$. Moreover, even for AGN hosting galaxies (e.g. X-ray or IRAC-selected AGN), their rest-optical light is still predominantly from stars. The impact of AGN, if not accounted for during morphological fitting, can only affect the size measurement on the order of $\approx10\%$ (see Section 3.2.2 and Figure 7 in \citealt{Ji2022AGN}), which is insufficient to explain the very large $\Sigma_e$ obtained here. Therefore, we argue that the very large $\Sigma_e$ found here are most likely due to the very compact stellar-mass distributions in the $z>3$ quiescent galaxies.

The strong correlation between the mass surface density and star-formation properties found at low redshifts has long been suggested as an indication of the causal link between the buildup of central regions of galaxies and quenching \citep[][just to name a few]{Kauffmann2003,Franx2008,Cheung2012,Barro2017,Lee2018,Ji2023}. However, because the most quiescent galaxies observed at $z<3$ are already old, with a typical stellar age of $>1$ Gyr \citep{Carnall2019b,Tacchella2022,Ji2022}, such a timescale is comparable to/longer than some post-quenching processes required to alter galaxy structures \citep[e.g., see Figure 10 in][]{Tacconi2020}. This makes it difficult to definitively answer whether the large mass surface densities found in quiescent galaxies are the cause or the consequence of quenching, or a by-product of an as-yet unidentified physical process.  Compared to the lower-redshift ones, the $z>3$ quiescent galaxies presented here are much younger, with stellar ages ranging from 100$-$700 Myr and a typical value of 500 Myr. These freshly quenched galaxies are already compact, suggesting the possible coeval formation of dense central regions and quenching at $z>3$ which is consistent with a recent JWST study of $3<z<4.5$ galaxies with forming stellar cores \citep{Ji2023jems}.

\subsubsection{GS-9209 at $z_{spec}=4.658$} \label{dis:9209}

The most remarkable $z>3$ quiescent galaxy in our sample is GS-9209, one of the earliest massive quiescent galaxy known so far having spectroscopically confirmed $z_{\rm{spec}}=4.658$ with JWST/NIRSpec observations \citep{Carnall2023}. The NIRSpec spectrum of GS-9209 is very similar to typical post-starburst galaxies, showing prominent stellar absorption features of A-type stars. Despite that AGN is not needed to fit GS-9209's broad-band SED, a broad H$\alpha$ emission line with FWHM $\sim 10^4$ km/s is observed in the NIRSpec spectrum, indicating that the galaxy actually hosts a faint AGN \citep{Carnall2023}. Moreover, as Figure  \ref{fig:Se_mass} shows, the galaxy has an extreme $\Sigma_e$, namely that it is comparable to the maximum allowed stellar mass surface density set by stellar feedback \citep{Grudic2019}. 
 
\begin{figure}
    \centering
    \includegraphics[width=0.47\textwidth]{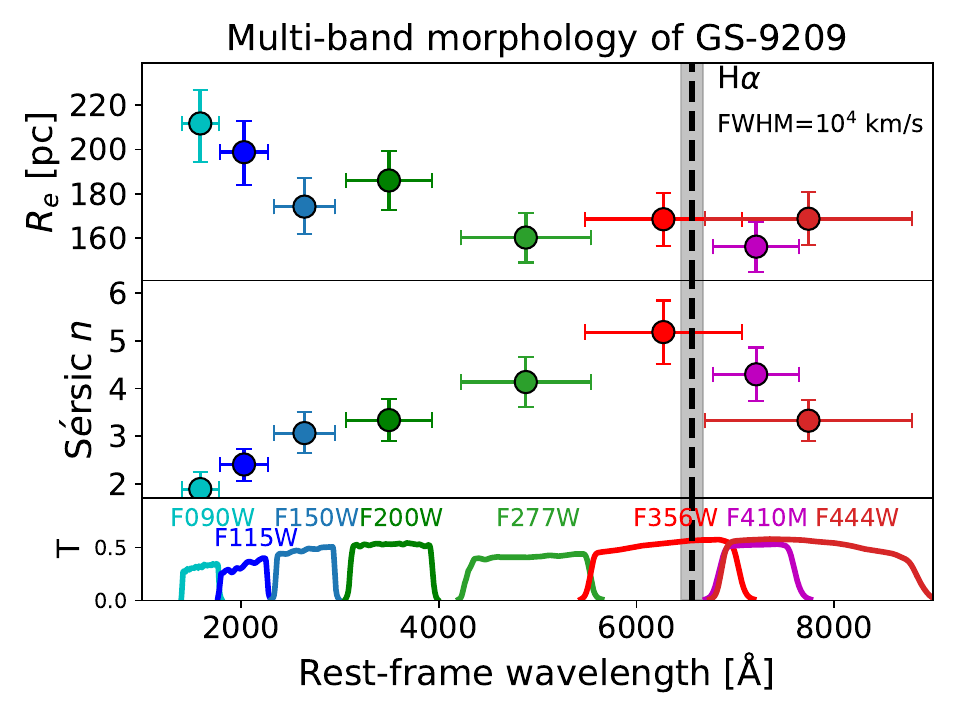}
    \caption{Multi-band morphologies of GS-9209 in NIRCam images from JADES. The black, vertical dashed line marks H$\alpha$ emission, and the shaded grey region corresponds to FWHM $=10,000$ km/s measured by \citet{Carnall2023}. Generally speaking, from rest-UV to NIR, GS-9209 becomes more compact (smaller $R_e$) and spheroidal (larger S\'{e}rsic $n$). Evidence of broad H$\alpha$ affecting the light distribution in F356W is also observed.}
    \label{fig:9209_morph}
\end{figure}

With deep NIRCam images from JADES, we measure the multi-band morphologies for GS-9209. The results are shown in Figure \ref{fig:9209_morph}. \citet{Carnall2023} reported a size of $215\pm20$ pc using the NIRCam F210M (rest-0.4$\micron$) image from JEMS \citep{Williams2023}, which is consistent with our measurement where we obtain the half-light semi-major radius of $223\pm16$ pc (or $186\pm13$ pc for circularized size $R_e$) in F200W. As the top panel of Figure \ref{fig:9209_morph} shows, overall the size of GS-9209 decreases with wavelength -- its light distribution becomes more compact at longer wavelength. The $R_e$ of the galaxy is $160\pm11$ pc in F277W (rest-0.5$\micron$) and $168\pm12$ pc in F444W (rest-1$\micron$), which are $2\sim3$ times smaller than the sizes predicted by using our best-fit size evolution (Section \ref{sec:size_evo}). In addition,  shown in the middle panel of Figure \ref{fig:9209_morph}, the S\'{e}rsic index $n$ also changes with wavelength. The morphology of GS-9209 gets more spheroidal at longer wavelengths. The best-fit S\'{e}rsic index is $n=1.8\pm0.4$ in F090W (rest-1500\AA), whereas it becomes $n=3.3\pm0.4$ in F444W (rest-1$\micron$). These immediately show the presence of radial color gradient in GS-9209. The galaxy is more extended in rest-UV and optical than in rest-NIR, either due to the radial gradient of dust attenuation where the central regions of the galaxy either have more dust, or older stellar populations, or both. 

Finally, we also note the sudden change of the light distribution of GS-9209 in F356W where the H$\alpha$ emission of the galaxy is covered. The galaxy is more nucleated (S\'{e}rsic $n=5.2\pm0.7$) in F356W than in adjacent filters, as the AGN H$\alpha$ emission comes from the very central parts of the galaxy. In the future, we may be able to recover some faint high-redshift AGN candidates missed by photometric selection methods through morphological analysis of deep NIRCam images, once the galaxies' redshifts are well constrained. 

\section{Caveats}
\label{sec:caveats}

We summarize here the main caveats of this work, particularly those associated with our conclusions regarding quiescent galaxies at $z>3$.

Our quiescent-galaxy sample is drawn from the JADES imaging in GOODS-South, which (at the time of this analysis) covers only $\sim 60~\mathrm{arcmin}^2$. Consequently, while the data are ultra-deep, the number ($\sim 10$) of massive quiescent galaxies at $z>3$ remains small. This limited sample size prevents us from drawing statistically significant conclusions about the size evolution of quiescent galaxies at $z>3$. However, since there is currently a lack of systematic environmental analyses for $z>3$ quiescent galaxies, and there is no indication that the $z>3$ quiescent galaxies in GOODS-S reside in large-scale structures (e.g., overdensities) that differ significantly from the underlying quiescent population at these redshifts, it is reasonable to regard the $z>3$ objects in our sample as a randomly drawn, small subsample of the underlying quiescent population. We therefore consider them to provide meaningful, if limited, insights into the structural properties of quiescent galaxies at these early epochs.

We remind the reader that our size-evolution fits are performed on individual galaxies (Section~\ref{sec:size_evo}), and the sample is dominated by lower-redshift objects (most lie at $z\lesssim 2$), which therefore drive the best-fit size-evolution parameters. We report the empirical observation that the $z>3$ quiescent galaxies in our sample appear very compact and may lie below an extrapolation of the lower-redshift size--redshift relation. However, we stress that we do \emph{not} claim that the current dataset can statistically establish whether the slope of size evolution changes at $z>3$; doing so will require substantially larger $z>3$ quiescent samples from wider-area JWST imaging. 

Similarly, we emphasize that the best-fit size-mass relation obtained in Section~\ref{sec:mass-size} should be used with caution at $z>3$ given the very limited sample size.

At $z>3$, many quiescent galaxies are very compact and approach the regime where the PSF, the degeneracy between $R_{\rm e}$ and S\'ersic index $n$, and potential mismatches between real light profiles and single-S\'ersic models can impact parametric size measurements. To assess the robustness of our conclusions, we compared our fiducial GALFIT sizes to multiple alternative approaches (Lenstronomy, fixed-$n$ fits, and a nonparametric Richardson-Lucy deconvolution-based estimate), finding that the different methods correlate well and that the relative size differences are typically $\lesssim 50\%$ for the majority of the $z>3$ objects. Nevertheless, for the most extreme systems, GALFIT can converge to the imposed upper bound in $n$ (and/or yield very compact cores), in which case parametric size estimates become less reliable. We treat such cases conservatively in our analysis and rely on alternative size estimates where appropriate. Future studies that characterize the detailed light distributions of $z>3$ quiescent galaxy populations (e.g., using multi-component modeling and nonparametric structural diagnostics), together with realistic image simulations that forward-model compact galaxies through the PSF, noise, and fitting pipelines, will be essential to quantify and correct systematic uncertainties in size measurements for these extremely compact systems.

\section{Summary}

In this work, we studied the size evolution of massive quiescent galaxies with $\log (M_*/M_\sun)>10$ at $0.5<z<5$. With multi-band deep NIRCam imaging from the JADES survey in the GOODS-South field, we characterized the size evolution of massive quiescent galaxies from UV through NIR at three different wavelengths, namely rest-frame 0.3, 0.5 and 1$\micron$. We found that 
\begin{enumerate}
	\item Regardless of wavelengths, the size of massive quiescent galaxies strongly evolves with cosmic time following $R_e\propto(1+z)^{-1.3}$. The pace of size evolution depends on stellar mass, with the evolution being faster for more massive galaxies. Specifically,  the size evolution follows $R_e\propto(1+z)^{-1.1}$ for lower-mass quiescent galaxies with $10<\log (M_*/M_\sun)<10.6$ , while it follows $R_e\propto(1+z)^{-1.7}$ for higher-mass quiescent galaxies with $\log (M_*/M_\sun)>10.6$ (Section \ref{sec:pace}). 
	\item As a result of negative stellar mass-to-light ratio, the size of quiescent galaxies decreases with wavelength. Relative to their rest-1$\micron$ sizes, on average quiescent galaxies are larger by 45\% at rest-0.3$\micron$ and 15\% at rest-0.5$\micron$, showing that the stellar mass distribution of quiescent galaxies is more compact than the rest-UV and optical light distributions. Evidence that younger quiescent galaxies have shallower color gradients than older ones is also seen, but requires future larger quiescent samples to test (Section \ref{sec:interc}). 
	\item At a fixed stellar mass, the size of quiescent galaxies is significantly larger at lower redshifts. The slope of the relationship between stellar mass and size only very weakly, if at all, depends on redshift, although we stress that the current sample size of $z>3$ quiescent galaxies is very limiting and future larger samples are needed to check if the slope remains unchanged towards higher redshifts (Section \ref{sec:mass-size}).
\end{enumerate}

We discussed possible physical mechanisms responsible for the observed size evolution. In particular, we studied the relationship between $R_e$ and $z_{\rm{form}}$, the formation redshift of galaxies. We showed that
\begin{enumerate}
	\item For lower-mass ($10<\log (M_*/M_\sun)<10.6$) quiescent galaxies, the slope of the relationship between $R_e$ and $z_{\rm{form}}$ is consistent with $-1$, the prediction of the progenitor effect. Thus we think that the progenitor effect may play a significant role in driving the apparent size evolution of quiescent galaxies of this stellar-mass range (Section \ref{sec:dis_size_evo_lowM}). 
	\item For higher-mass ($\log (M_*/M_\sun)>10.6$) quiescent galaxies, the relationship between $R_e$ and $z_{\rm{form}}$ depends on stellar age. For young quiescent galaxies, the slope of the $R_e$-$z_{\rm{form}}$ relationship is consistent with $-1$ within uncertainty (though somewhat steeper), whereas for old quiescent galaxies the relationship is much steeper than the prediction of the progenitor effect alone. These are in line with the picture that, in addition to the progenitor effect, mergers and/or post-quenching continuous gas accretion drive additional, apparent size growth in very massive quiescent galaxies (Section \ref{sec:dis_size_evo_highM}).
\end{enumerate}

Notwithstanding the very limited sample size, we studied the structure of massive quiescent galaxies at $z>3$ in this work. We saw evidence that these quiescent galaxies are even smaller than anticipated using the best-fit size evolution derived for lower-redshift quiescent galaxies (Section \ref{sec:pace}). Moreover, we showed that these freshly quenched quiescent galaxies, with a typical stellar age of 500 Myr compared to $>1$ Gyr observed in quiescent galaxies at lower redshifts, are already very compact, with $\Sigma_e \gtrsim 10^{10} M_\sun/\rm{kpc}^2$, suggesting the possible coeval formation of dense central regions and quenching at $z>3$ (Section \ref{diss:zgt3}). We also studied in detail the multi-band morphology for GS-9209 \citep{Carnall2023}, one of the earliest massive quiescent galaxies known so far with $z_{\rm{spec}} = 4.658$. From rest-UV to NIR, we found that GS-9209 becomes increasingly smaller and more compact, and its light profile becomes more spheroidal, showing that the radial color gradient is already present in this earliest massive quiescent galaxy (Section \ref{dis:9209}).

Finally, as a closing remark, we note that the current high-redshift quiescent sample, e.g. at $z>3$, is still small. Explicitly and better constraining the size evolution at $z>3$ will await larger samples selected from JWST imaging, and this motivates future analysis based on robustly JWST-selected high-redshift samples \citep[e.g.,][]{Alberts2023} over larger sky areas. We also stress that characterizing the size evolution really is just the first step to constrain the formation pathways of the first quiescent galaxies in the Universe. The structure of the earliest quiescent galaxies is likely much more complex. Some hints have already been seen from our Figure \ref{fig:size_zgt3}: Sizes from the Richardson–Lucy deconvolution approach are larger than those from the single S\'{e}rsic fitting, suggesting multi-component structures are already present in quiescent galaxies at $z>3$. With the superb NIR imaging from JWST, we will be able to study such complex structural properties in the earliest quiescent galaxies in unprecedented detail.

\begin{acknowledgments}
ZJ, BDJ, BER, ZC, CD, DJE, EE, KH, JMH, GR, MR, FS and CNAW acknowledge support from JWST/NIRCam contract to the University of Arizona, NAS5-02015. 
The research of CCW is supported by NOIRLab, which is managed by the Association of Universities for Research in Astronomy (AURA) under a cooperative agreement with the National Science Foundation.
BER acknowledges support from JWST Program 3215. 
SA acknowledges support from the JWST Mid-Infrared Instrument (MIRI) Science Team Lead, grant 80NSSC18K0555, from NASA Goddard Space Flight Center to the University of Arizona.
WB, FDE, TJL, RM, LS and JW acknowledge support by the Science and Technology Facilities Council (STFC), by the ERC through Advanced Grant 695671 ``QUENCH''. 
KB acknowledges support in part by the Australian Research Council Centre of Excellence for All Sky Astrophysics in 3 Dimensions (ASTRO 3D), through project number CE170100013.
AJB and JC acknowledge funding from the ``FirstGalaxies'' Advanced Grant from the European Research Council (ERC) under the European Union's Horizon 2020 research and innovation programme (Grant agreement No. 789056). 
SC acknowledges support by European Union's HE ERC Starting Grant No. 101040227 - WINGS.
ECL acknowledges support of an STFC Webb Fellowship (ST/W001438/1).
DJE is supported as a Simons Investigator. 
RH acknowledges that funding for this research was provided by the Johns Hopkins University, Institute for Data Intensive Engineering and Science (IDIES).
RM acknowledges support by the UKRI Frontier Research grant RISEandFALL, and funding from a research professorship from the Royal Society.
H{\"U} gratefully acknowledges support by the Isaac Newton Trust and by the Kavli Foundation through a Newton-Kavli Junior Fellowship. H{\"U} acknowledges funding by the European Union (ERC APEX, 101164796). Views and opinions expressed are however those of the authors only and do not necessarily reflect those of the European Union or the European Research Council Executive Agency. Neither the European Union nor the granting authority can be held responsible for them.

This work made use of the {\it lux} supercomputer at UC Santa Cruz which is funded by NSF MRI grant AST 1828315, as well as the High Performance Computing (HPC) resources at the University
of Arizona which is funded by the Office of Research Discovery and Innovation (ORDI), Chief Information Officer (CIO), and University Information Technology Services (UITS).
\end{acknowledgments}

\facilities{HST (ACS, WFC3), JWST (NIRCam)}
\software{{\sc Galfit} \citep{galfit}, \prospector \citep{Johnson2021}, {\sc FSPS} \citep{Conroy2009,Conroy2010}, {\sc MIST} \citep{Choi2016,Dotter2016}, {\sc MILES} \citep{Falcon-Barroso2011}, {\sc astropy} \citep{astropy}, {\sc photutils} \citep{photutils}, {\sc Webbpsf} \citep{Perrin2014}, {\sc numpy} \citep{numpy}, {\sc scipy} \citep{scipy}, {\sc corner} \citep{corner}}

\appendix

\section{SED fitting with different SFHs } \label{app:prior}
\begin{figure*}
    \includegraphics[width=0.97\textwidth]{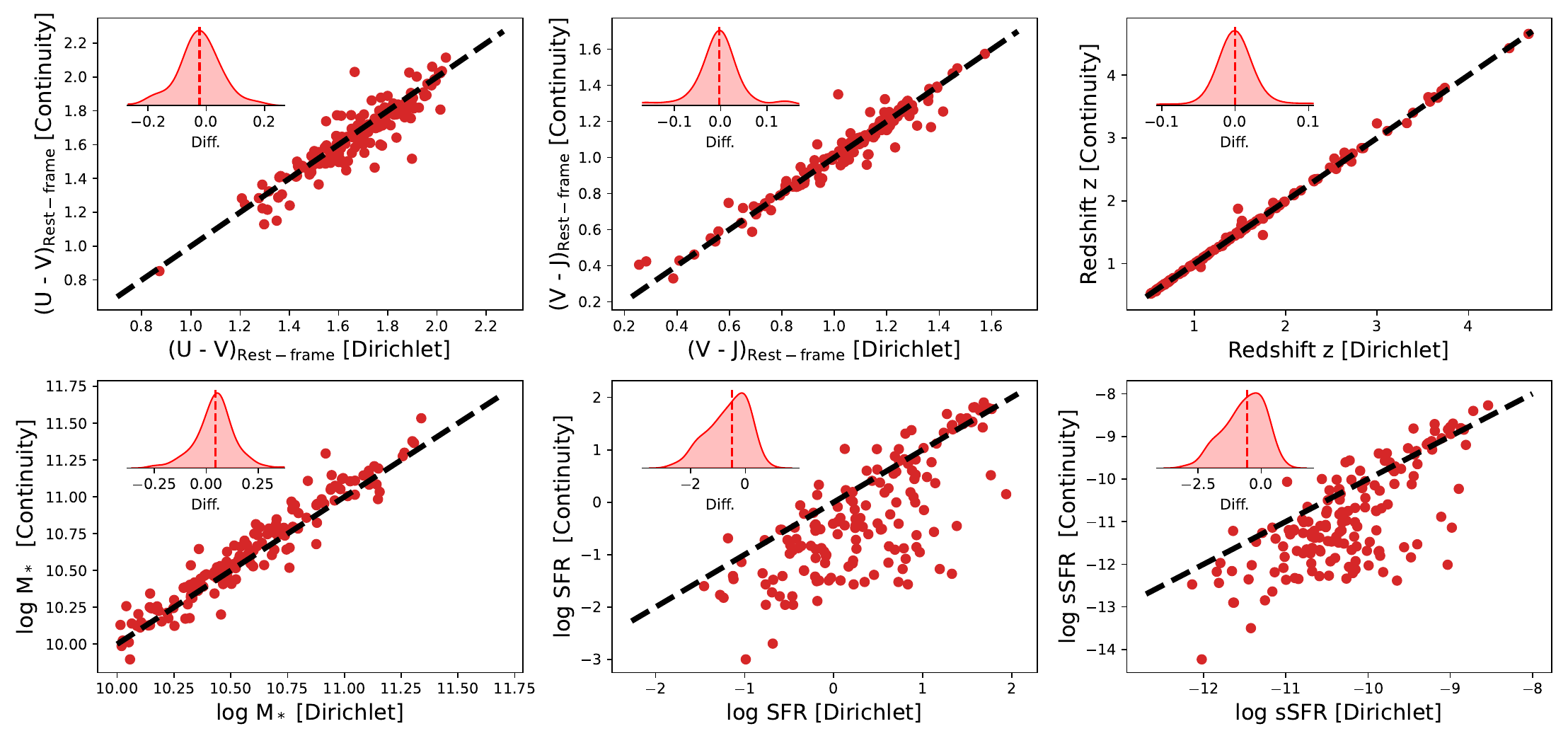}
    \caption{Comparison of physical properties derived from nonparametric SFHs with the Dirichlet prior ($x$-axis) and nonparametric SFHs with the continuity prior ($y$-axis).}
    \label{fig:dir_con}
\end{figure*}

\begin{figure*}
    \includegraphics[width=0.97\textwidth]{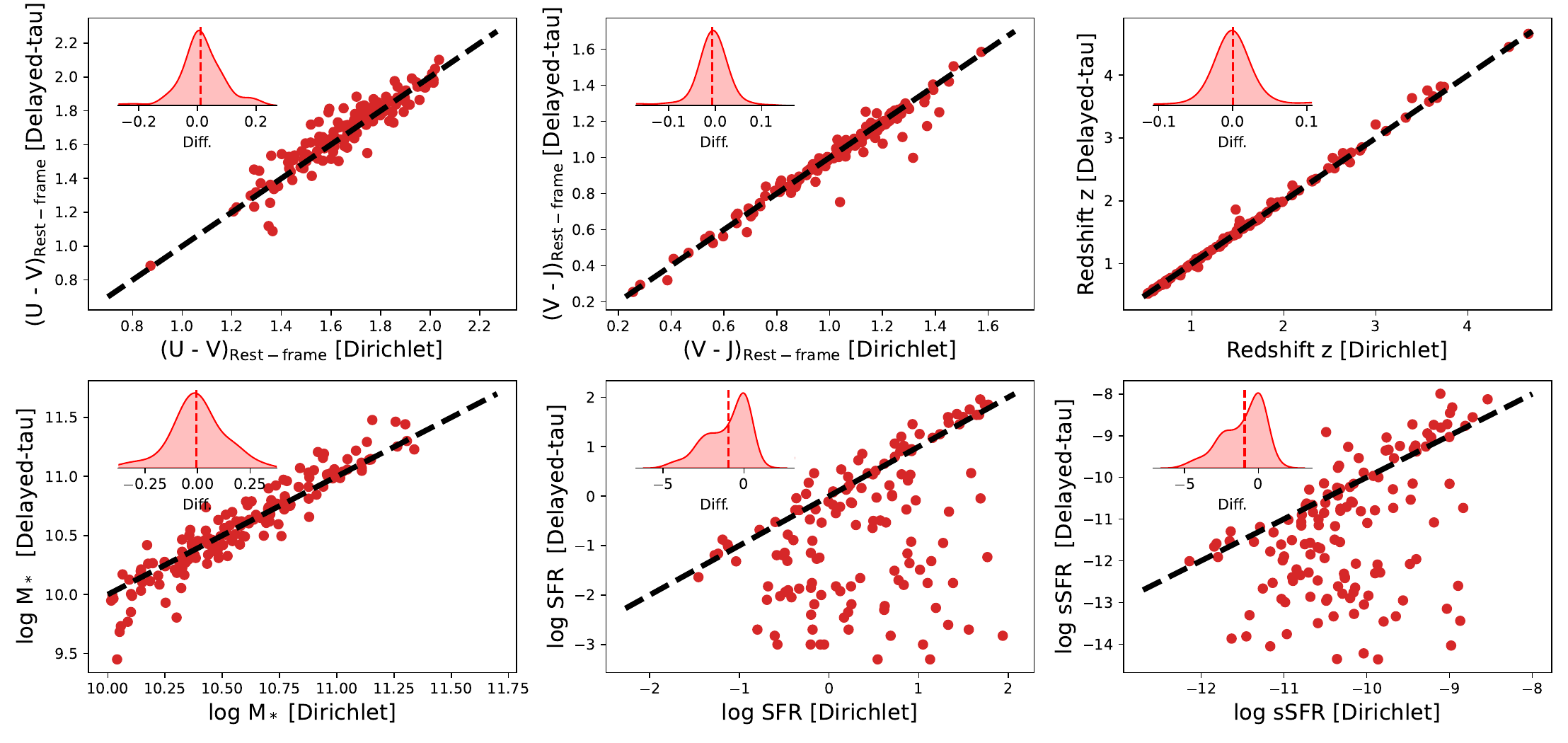}
    \caption{Similar to Figure \ref{fig:dir_con}, but now we compare the physical properties from nonparametric Dirichlet SFHs with those from parametric delayed-tau SFHs.}
    \label{fig:dir_param}
\end{figure*}

We test our SED fitting with different assumptions of SFHs. Specifically, in Figure \ref{fig:dir_con}, we compare the physical quantities derived using our fiducial SED assumptions, i.e. nonparametric SFHs with the Dirichlet prior (Section \ref{sec:sed}), with those derived using nonparametric SFHs with the continuity prior. These two nonparametric SFH priors are very similar, except that the continuity prior disfavor sudden changes of SFRs in adjacent lookback time bins (see \citealt{Leja2019} for details). Similarly, in Figure \ref{fig:dir_param}, we compare the our fiducial SED fitting results with those derived using parametric delayed-tau SFHs. Stellar masses, redshifts and rest-frame UVJ colors agree very well among different SFH assumptions. Systematic offsets exist for the SFR measures for quiescent galaxies, as expected for SED fitting with photometric data only \citep{Conroy2013}. 

\bibliography{ji_2023_jades_qg_size}{}
\bibliographystyle{aasjournal}

\end{document}